\documentclass[11pt]{article}

\newcommand{\pt}{\noindent {$\bullet$~}}

\def\tr{\;{\rm tr}\;}

\def\bra{\langle}   \def\ket{\rangle}
\def\pr{\prime}
\def\vs#1{\vspace{#1mm}}
\def\implies{\Rightarrow}
\def\fl{\flushleft}

\newcommand{\tl}[1]{\tilde{#1}}

\newcommand{\dd}[2]{\frac {\partial #1}{\partial #2}}
\newcommand{\pdr}{\partial}

\newcommand{\beq}{\begin{eqnarray}}
\newcommand{\eeq}{\end{eqnarray}}

\newcommand{\un}[1]{\underline{#1}}

\newcommand{\half}{\frac{1}{2}}

\newcommand{\ov}[1]{\frac{1}{#1}}

\def\a{\alpha}                
\def\d{\delta}        \def\eps{\epsilon} 
        \def\ka{\kappa} \def\la{\lambda}      \def\La{\Lambda}
             \def\F{\Phi}         
             
\def\r{\varrho}       \def\SS{\Sigma}

\newcommand{\N}{{1 \over N}}
\newcommand{\Nsq}{{1 \over N^2}}
\newcommand{\Ntr}{{{\rm tr} \over N}}

\usepackage{hyperref}
\usepackage{graphics}
\usepackage{graphicx}
\usepackage{epsf}

\begin{document}

%------------------------------------------------

\begin{titlepage}

% \vs{-40}

\title{\normalsize \hfill ITP-UU-06/49, SPIN-06/39
\\ \hfill {\tt hep-th/0611350}\\ \vskip 0mm \Large\bf
Non-anomalous `Ward' identities to supplement large-$N$ multi-matrix
loop equations for correlations}

\author{Levent Akant$^a$ and Govind S. Krishnaswami$^b$}
\date{\normalsize $^a$ Feza G\"{u}rsey Institute, \\
Emek Mah.Rasathane Yolu  No:68, 34684, \\ \c{C}engelk\"{o}y,
Istanbul, Turkey \\ \smallskip and \\ 
$^b$ Institute for Theoretical Physics \& Spinoza Institute \\
Utrecht University, Postbus 80.195, 3508 TD, Utrecht, The Netherlands
\smallskip \\ e-mail: \tt akant@gursey.gov.tr, ~~g.s.krishnaswami@phys.uu.nl \\ \medskip
November 30, 2006}

\maketitle

\begin{quotation} \noindent {\large\bf Abstract }

{

This work concerns single-trace correlations of Euclidean
multi-matrix models. In the large-$N$ limit we show that
Schwinger-Dyson equations (SDE) imply loop equations (LE) and
non-anomalous Ward identities (WI). LE are associated to
generic infinitesimal changes of matrix variables (vector
fields). WI correspond to vector fields preserving measure and
action. The former are analogous to Makeenko-Migdal equations
and the latter to Slavnov-Taylor identities. LE correspond to
leading large-$N$ SDE. WI correspond to $1/N^2$ suppressed SDE.
But they become leading equations since LE for non-anomalous
vector fields are vacuous. We show that symmetries at
$N=\infty$ persist at finite $N$, preventing mixing with
multi-trace correlations. For $1$ matrix, there are no
non-anomalous infinitesimal symmetries. For $2$ or more
matrices, measure preserving vector fields form an infinite
dimensional graded Lie algebra, and non-anomalous action preserving ones a
subalgebra. For Gaussian, Chern-Simons and Yang-Mills models we
identify up to cubic non-anomalous vector fields, though they
can be arbitrarily non-linear. WI are homogeneous linear
equations. We use them with the LE to determine some
correlations of these models. WI alleviate the underdeterminacy
of LE. Non-anomalous symmetries give a naturalness-type
explanation for why several linear combinations of correlations
in these models vanish.

}

\end{quotation}

\vfill \flushleft



% MSC: 16W30, 16W25, 16S80, 20E05, 81T13

 Journal reference: 	JHEP 0702 (2007) 073

\thispagestyle{empty}

\end{titlepage}

\eject
%--------------------------------------------------

\scriptsize

{\tableofcontents}

\normalsize

\clearpage

%----------------------------------------------------------
\section{Introduction}
\label{s-intro}
%----------------------------------------------------------

Hermitian multi-matrix models are quantum systems where the
dynamical variables are a set of $N \times N$ hermitian
matrices. Observables must be basis independent, i.e. invariant
under the global adjoint action of $U(N)$ on the matrices.
Expectation values of observables are determined by an average
over all matrix elements with respect to a Boltzmann weight
specified by an action. Matrix models simplify in 't Hooft's
large-$N$ limit, since fluctuations in $U(N)$-invariant
observables are small in this limit.

Multi-matrix models are simplified models for the dynamics of
gauge fields in Yang-Mills theory. It is a fundamental and
challenging problem to determine the free energy and
correlation functions of multi-matrix models, and elucidate the
mathematical framework needed to study them. Multi-matrix
models are much harder to understand than single-matrix models,
but also have a much richer structure. Large-$N$ matrix models
and more generally large-$N$ gauge theories have been studied
ever since their relevance as an approximation to the theory of
strong interactions was pointed out by 't Hooft in the mid
1970s \cite{thooft-large-N,bipz,witten-baryons-N}. Important
progress in obtaining the loop equations of Yang-Mills theory
and study of the large-$N$ limit was made in the late 1970s and
early 1980s by Migdal and Makeenko \cite{makeenko-migdal-eqn},
Cvitanovic\cite{Cvitanovic:1980jz}, Yaffe\cite{Yaffe}, Jevicki
and Sakita\cite{Jevicki:1979mb} and others \cite{Wadia:1980rb}.
The subject was applied to random surface theory, 2d string
theory and the matrix approach to M-theory in the 1990s.
Meanwhile, there has been a steady stream of developments in
matrix models of which we cite a few examples. These include
their connections to non-commutative probability theory
\cite{voiculescu,douglas,gopakumar-gross,douglas-li,entropy-var-ppl,hamiltonian-mat-mod-fisher-info},
the study of multi-matrix symmetry algebras and their
connections to spin chains
\cite{lee-rajeev-closed,Minahan:2002ve}, exact
solutions\cite{kazakov-marshakov} and their relation to
CFT\cite{Kostov:1999xi} and algebraic geometry and detailed
studies of the loop
equations\cite{eynard-loop-eqn-chain,gsk-loop-eqns}. Much of
the existing literature deals with 1-matrix models or exact
solutions for specific observables of carefully chosen
multi-matrix models. We hope to complement this by developing a
framework and methods that apply to general multi-matrix
models.

Throughout physics, we exploit symmetries to simplify dynamical
equations by reducing the number of unknowns. The quantum
dynamical equations of a large-$N$ multi-matrix model are the
loop equations\footnote{The name loop equations is used because
these equations are analogous to the Makeenko-Migdal equations
of Yang-Mills theory, which were formulated for the Wilson
loops. The word loop has nothing to do with loops in Feynman
diagrams. Another name for these equations is factorized
Schwinger-Dyson equations.} for single-trace correlations.
These correlations are analogs of gluon and ghost correlation
functions of Yang-Mills theory. Here, we develop a general
framework to find non-anomalous infinitesimal symmetries of
multi-matrix models (i.e. those that preserve both action and
measure). These symmetries are used to infer Ward identities,
which supplement the loop equations to determine correlations.
These non-anomalous symmetries and Ward identities can be
regarded as finite dimensional analogues of BRST invariance and
Slavnov-Taylor identities of Yang-Mills theory. The ideas are
illustrated with examples from $2$ and $3$-matrix models.

To motivate this work, we explain how we came to think along
these lines. We were trying to solve the loop equations (LE) to
determine large-$N$ single trace correlations of some specific
multi-matrix models\cite{gsk-loop-eqns}. Single-trace
correlations are the basic objects of interest, since
multi-trace correlations factorize into products of
single-trace ones in the large-$N$ limit. The LE state the
invariance of the partition function under infinitesimal but
non-linear changes of integration variables, in the large-$N$
limit. They relate a change in action to a change in measure.
Such infinitesimal changes of integration variables can be
regarded as vector fields. A priori, there is one LE for each
such vector field. In many non-Gaussian cases, we found the LE
were underdetermined. Moreover, this underdeterminacy seemed
related to the fact that for several changes of variable, the
LE were vacuous. In other words, for some vector fields, both
the change in action and change in measure simultaneously
vanished in the large-$N$ limit. We were looking for additional
equations to supplement the LE and alleviate their
underdeterminacy.

Now, the LE can be regarded as the large-$N$ limit of the
finite-$N$ Schwinger-Dyson equations (SDE). The SDE are
conditions for the invariance of matrix integrals for
multi-trace correlations under infinitesimal non-linear changes
of integration variables. They relate a change in action to a
sum of a change in measure and change in observable. While the
first two are usually of order $N^0$, the latter\footnote{There
is no ${\cal O}(1/N)$ contribution in a matrix model, i.e. in
the absence of quarks or $N$-vectors.} is usually of order
$N^{-2}$. So naively, in the large-$N$ limit, the latter drops
out and using factorization, we get back the LE\footnote{This
is why LE are also called factorized Schwinger-Dyson equations.
The name Virasoro constraints is also used, especially in
applications to string theory.}. However, in the special case
where the vector field is a symmetry of both action and measure
in the large-$N$ limit, the naive large-$N$ limit of the SDE is
vacuous and one must go to the next order in $1/N^2$ to get the
leading condition. For generic vector fields, this ${\cal
O}(1/N^2)$ SDE would not be an equation for the single-trace
correlations alone, since it involves ${\cal O}(1/N^2)$
corrections to the factorized result for multi-trace
correlations. However, remarkably, we show that if a vector
field is a symmetry of the action in the large-$N$ limit, then
it is also a symmetry\footnote{There are potentially more
symmetries at finite-$N$ than at large-$N$, so the converse is
almost certainly false.} at each order in $1/N^2$. The same
also holds for symmetries of measure. The simple reason is that
there are more independent variables as $N$ grows, and so more
conditions on a vector field to be a symmetry as $N$ increases.
Non-anomalous vector fields define simultaneous symmetries of
action and measure in the large-$N$ limit. Thus, for
non-anomalous vector fields, the change in action and measure
terms drop out of the ${\cal O}(1/N^2)$ SDE, which then becomes
a condition for invariance of the expectation value of the
observable in the large-$N$ limit, schematically $L_v G_{i_1
\cdots i_n} =0$. This latter condition only involves
single-trace correlations $G_{i_1 \cdots i_n}$, and is what we
call a non-anomalous Ward identity (WI). WI are associated to
vector fields $v$ that leave both action and measure invariant
in the large-$N$ limit. $L_v$ is the Lie derivative along $v$.
It is precisely for such vector fields that the LE are vacuous.
We show that such vector fields form an infinite dimensional
graded Lie algebra. Thus, as $N \to \infty$, the SDE imply not
just the LE, which are associated to generic vector fields, but
also WI, which are associated to non-anomalous vector fields.
The latter are easily overlooked in a naive passage to the
large-$N$ limit. Naively, the WI appear to be `universal', i.e.
independent of the matrix model action, but this is not really
true. Whether or not a WI holds is determined by whether the
corresponding vector field is a simultaneous symmetry of both
action and measure.

In retrospect, these non-anomalous symmetries and WI are not
unexpected. In Yang-Mills theory, non-anomalous symmetries
include Poincare and BRST invariance. WI for the latter are
Slavnov-Taylor identities. Our non-anomalous WI share with the
Slavnov-Taylor identities the structural similarity of being
homogeneous linear equations for correlations. Just like our
WI, the Slavnov-Taylor identities seem independent of the
gauge-fixed Yang-Mills action, until one realizes they hold
only because the action and measure are BRST invariant. In
Yang-Mills theory, while Poincare transformations act linearly
on the fields, BRST transformations are quadratically
non-linear. For specific matrix models, we find non-anomalous
symmetries that are linear, quadratic $(n=2)$ and cubic
$(n=3)$; there is no limit to the possible non-linearity of
such symmetries. Moreover, for $n>1$, some rank-$n+1$
non-anomalous symmetries can be obtained via the Lie brackets
of rank-$n$ symmetries. This is reminiscent of how Poisson
brackets of conserved charges (if non-vanishing), give higher
conserved charges in integrable models.

We find a significant difference between $1$-matrix models and
multi-matrix models. The measure for a single matrix in the
large-$N$ limit admits only one continuous symmetry, i.e.
translations of the matrix. Translations, however, are not a
symmetry of any non-trivial $1$-matrix action. Thus,
non-trivial $1$-matrix models have no non-anomalous WI.
Interestingly, we find that the measures for multi-matrix
models allow for large classes of symmetries, some of which may
also be symmetries of a given action.

In practice, once a non-anomalous symmetry of a model is known,
it is easier to first solve the resulting WI and then consider
the LE. The WI, being homogeneous linear equations, force
several correlations or linear combinations thereof to vanish.
This simplifies analysis of the LE, which are mildly
non-linear. However, we caution that some WI may contain the
same information as contained in the LE, while others may
provide new conditions.

We emphasize that the techniques and results of this paper are
exact. They do not involve any approximation beyond the passage
to $N=\infty$. Our methods apply to single-trace correlations
of generic hermitian multi-matrix models with polynomial
actions. They are not special to any subclass of actions or
correlations. Of course, the non-anomalous symmetries and WI
will depend on which model we consider. One lesson we learned
is that though it is a bit laborious, it is possible to solve
the LE and WI of large-$N$ multi-matrix models to determine
exact correlations, starting from the lowest rank ones.

Finally, our derivation of the SDE, LE and WI makes use of the
matrix integral representation for correlations. In cases where
these integrals converge, we expect the equations to be
rigorously valid. When the matrix integrals diverge, the SDE,
LE and WI are only formal statements and their consistency is
not guaranteed by our work. Indeed, we seem to find an example
where formal use of these equations for a model whose matrix
integrals diverge, leads to inconsistencies. We are yet to
understand the deeper significance of this.

We can give another interpretation of our results. Suppose one
were to calculate single trace correlations of a large-$N$
multi-matrix model. Then in many cases one would find there are
several linear combinations of correlations that vanish. One
might look for a naturalness-type explanation for this i.e., a
non-anomalous symmetry that forces those linear combinations to
vanish. In some cases, there is a discrete symmetry (such as $A
\to -A$ for correlations of odd order for an even action) that
does the job. The results of this paper may be regarded as the
discovery of several new continuous non-anomalous symmetries of
multi-matrix models. For example, $\d A_1 = a[A_1,[A_2,A_1]]$
and $\d A_2 = a[[A_1,A_2],A_2]$ is a non-anomalous symmetry of
the Gaussian+YM 2-matrix model for all real $a$. Such
symmetries lead to WI, which ensure that the quantities in
question vanish.

{\bf Organization and summary of results:} In section
\ref{s-sde-and-ward-id} we determine the SDE of hermitian
multi-matrix models. We show that in the large-$N$ limit, they
lead to LE supplemented by WI. It is asserted that the WI are
to be imposed for every vector field that is a simultaneous
symmetry of action and measure in the large-$N$ limit. The
proof of validity of the WI is completed in sections
\ref{s-vol-pres-vfld-ann-terms-in-LE-and-SDE} and
\ref{s-proof-of-ward-id}. In section \ref{s-single-matrix} we
explain why the WI trivialize for a $1$-matrix model. Section
\ref{s-underdeterminacy-of-LE} exhibits that multi-matrix LE
are often underdetermined and this motivates the need for
additional equations to determine correlations. WI potentially
alleviate the underdeterminacy of LE. In section
\ref{s-meas-pres-transformations} we characterize measure
preserving vector fields of multi-matrix models in the
large-$N$ limit. We show that they form an infinite dimensional
Lie algebra (section \ref{s-vol-pres-vfld-form-lie-alg}).
Measure preserving transformations of $2$- and $3$-matrix
models are given in section
\ref{s-eg-meas-pres-vfld-2-3-mat-mod}. We work out the linear
and quadratic non-anomalous symmetries of Gaussian,
Chern-Simons, Yang-Mills and Gaussian+Yang-Mills multi-matrix
models in sections \ref{s-linear-symm} and
\ref{s-non-linear-symm} and also construct some cubic
symmetries via Lie brackets of quadratic ones. In section
\ref{s-ward-id} we explicitly give the LE and non-anomalous WI
for the Gaussian, Gaussian+YM and Chern-Simons models. We show
that several correlations vanish, determine some non-vanishing
correlations, and also obtain non-trivial relations among other
non-vanishing correlations. In section \ref{s-WI-ym} we show
that formal use of LE and WI for a model whose matrix integrals
do not converge potentially leads to inconsistencies. Some
outstanding questions are collected in \ref{s-discussion}. In
appendix \ref{a-hermitian-derivation-of-SDE} we give an
alternate derivation of the SDE that preserves hermiticity of
matrices. In appendix \ref{a-other-changes-of-var} we consider
some other possible changes of variables in an unsuccessful
search for equations satisfied by the $N=\infty$ single trace
correlations, over and above the LE and WI. In appendix
\ref{a-all-G_I-from-WI} we argue that the WI by themselves
(without use of LE) cannot determine all correlations of a
non-trivial model. In appendix \ref{a-cyclic-tensors-rank-n} we
quote a useful formula for the number of cyclically symmetric
tensors of rank-$n$ in a $\La$-matrix model.

%----------------------------------------------------------
\section{Schwinger-Dyson equations and Ward identities}
\label{s-sde-and-ward-id}
%----------------------------------------------------------

We consider a bosonic\footnote{It is possible to extend these
methods to models with gluon and ghost fields, such as gauge
fixed Yang-Mills theory. In this case the $A_i$ would include
hermitian matrices as well as matrices with grassmann entries.}
Euclidean matrix model with $\La$ random hermitian matrices
$A_i, i=1,2,\cdots, \La$. The action $\tr S(A) = \tr \sum_{|J|
\leq m} S^J A_J$ is taken to be a polynomial. Due to the trace,
only the cyclic projections of the coupling tensors $S^I$
contribute to the action. Multi-indices are denoted by capital
letters, for example, $I=i_1 i_2 \cdots i_n$. Repeated lower
and upper indices as summed and $|I|$ denotes the length of the
multi-index.

Observables are functions of $A_i$ that are invariant under the
global adjoint action $A_i \to U A_i U^\dag$ of $ U \in U(N)$.
An important class of such functions are the trace invariants
$\Phi_I = \N \tr A_I$. The partition function and multi-trace
correlations are defined as
    \beq
    Z = \int \prod_{j=1}^\La dA_j  e^{-N \tr S(A)} {\rm ~~and~~}
    \bra \Phi_{K_1} \cdots \Phi_{K_n} \ket = \ov{Z} \int \Pi_j dA_j  e^{-N \tr S(A)}
    \Phi_{K_1} \cdots \Phi_{K_n}.
    \label{e-def-of-partn-fn-correlations}
    \eeq
$\bra \Phi_{K_1} \cdots \Phi_{K_n} \ket$ is symmetric under
interchange of any pair from $K_1, \cdots, K_n$. It is
cyclically symmetric in each $K_i$ separately. $\bra \Phi_{K_1}
\cdots \Phi_{K_n} \ket$ may be expanded in inverse powers of
$N^2$
    \beq
    \bra \Phi_{K_1} \cdots \Phi_{K_n} \ket = G^{(0)}_{K_1; K_2; \cdots ; K_n}
    + \ov{N^2} G^{(2)}_{K_1 ; K_2 \cdots ; K_n} +
    \ov{N^4} G^{(4)}_{K_1 ; K_2 \cdots; K_n} + \cdots.
    \label{e-ovNsq-exp-of-corrlns}
    \eeq
The coefficient of $N^{-2h}$ can be regarded as a sum of
Feynman diagrams that can be drawn on a Riemann surface with
$h$ handles and $n$ disks cut out\footnote{A pictorial
representation of (\ref{e-ovNsq-exp-of-corrlns}) for a
four-point correlation would resemble Fig. $1.8$ on page $31$
of Ref. \cite{green-schwarz-witten}}. The perimeter of each
disk is associated to one of the inserted $K_i$'s. In
particular, for $h=0$, these are planar diagrams. Each of the
$G^{(2h)}_{K_1; \cdots ; K_n}$ is symmetric in the
multi-indices $K_1, \cdots, K_n$. Factorization of multi-trace
correlations in the large-$N$ limit\cite{makeenko-book} means
that $G^{(0)}$ can be written as a product of single trace
correlations
    \beq
    \lim_{N \to \infty} \bra \Phi_{K_1} \cdots \Phi_{K_n} \ket =
    G^{(0)}_{K_1; K_2; \cdots ; K_n} = G_{K_1} \cdots G_{K_n},
    {\rm ~~~where~~} G_K = \lim_{N \to \infty} \bra \Ntr \Phi_K
    \ket.
    \label{e-factorization}
    \eeq
The single-trace gluon correlations $G_K$ are cyclically
symmetric in $K$ and satisfy the hermiticity condition $G_K^* =
G_{\bar K}$ provided $S^I$ also satisfy this property. Here
$\bar K$ is the word $K$ with order of indices reversed. $G_K$
will also be referred to as moments, they are the moments of a
non-commutative probability distribution\cite{entropy-var-ppl}
when $\La > 1$. The rank of $G_K$ is defined as $|K|$.

To determine correlations, we derive Schwinger-Dyson equations
(SDE), conditions for invariance of matrix integrals for $\bra
\Phi_{K_1} \cdots \Phi_{K_n} \ket$ under infinitesimal
non-linear changes of variables\footnote{These changes of
variable don't always preserve hermiticity of $A_i$. Under a
change of integration variable, the integrand, measure as well
as domain of integration may change, but the value of the
integral is unchanged. It is possible to derive the SDE by
making hermitian changes of variable, see appendix
\ref{a-hermitian-derivation-of-SDE}.}
    \beq
    [A_i]^a_b \to [A^\pr_i]^a_b = [A_i]^a_b + v_i^I [A_I]^a_b, ~~~~
    v^I_i {\rm ~~ infinitesimal ~real~ parameters~for~} |I| \geq 0.
    \label{e-change-of-variable}
    \eeq
These include the BRST-type of transformations used to derive
the Slavnov-Taylor identities of gauge-fixed Yang-Mills theory.
For example, in Lorentz gauge the BRST transformations are
infinitesimal quadratic transformations ($\la$ is an
infinitesimal anti-commuting parameter),
    \beq
    A_\mu \to A_\mu + \la \pdr_\mu c + \la [A_\mu,c], ~~~
    c \to c + \la [c,c]_+, ~~~
    \bar c \to \bar c + \la \pdr^\mu A_\mu.
    \label{e-brst-transformation}
    \eeq
To calculate the effect of (\ref{e-change-of-variable}) on the
integral (\ref{e-def-of-partn-fn-correlations}) defining $\bra
\Phi_{K_1} \cdots \Phi_{K_n} \ket$ we need\footnote{$\bra
\Phi_{K_1} \cdots \Phi_{K_n} \ket$ is the quotient of two
integrals. Here we make a change of variable in the numerator
but not the integral for $Z$ in the denominator. We could
change variables in each, but this doesn't give new equations.}
the infinitesimal change in action, measure and the inserted
observable $\Phi_K$
    \beq
    e^{-N \tr S^J A_J} &\mapsto& e^{-N \tr S^J A_J}
        (1-N^2 v_i^I S^{J_1 i J_2} \Phi_{J_1 I J_2} ) + {\cal O}(v^2),
        \cr
    \det\bigg( \dd{[A^\pr_i]^a_b}{[A_j]^c_d} \bigg) &=&
        1 + N^2 v_i^I \d_I^{I_1 i I_2} \Phi_{I_1} \Phi_{I_2} + {\cal O}(v^2)\cr
    \Phi_K &\mapsto& \Phi_K + \d_K^{L i M} v_i^I
        \Phi_{L I M}.
    \eeq
The conditions for invariance of $\bra \Phi_{K_1} \cdots
\Phi_{K_n} \ket$ to linear order in $v_i^I$ are the finite $N$
SDE\footnote{For the purpose of deriving the loop equations and
Ward identities, it is adequate to start with SDE for single
trace correlations, so the reader could set $n=1$ in a first
reading.}
    \beq
    v_i^I S^{J_1 i J_2} \bra \F_{J_1 I J_2} \F_{K_1} \cdots \F_{K_n} \ket &=&
     v_i^I \d_I^{I_1 i I_2} \bra \Phi_{I_1} \F_{I_2} \F_{K_1} \cdots \F_{K_n} \ket
    \cr && + \ov{N^2} \sum_{p=1}^n \d_{K_p}^{L_p i M_p} v_i^I \bra \F_{K_1}
    \cdots \F_{K_{p-1}} \F_{L_p I M_p} \F_{K_{p+1}} \cdots \F_{K_n}
    \ket.
    \label{e-SDE}
    \eeq
There is a priori one such SDE for each vector field $v^I_i$
and each $n=0,1,2,\cdots$, where $n$ is the number of
insertions. The LHS is the expectation value of the change in
action (along with $\Phi_K$ insertions). The first term on the
RHS is the expectation value of the change in measure (with
$\F_K$ insertions) and the second term on the RHS is the
expectation value of the change in insertions $\Phi_K$. So far,
we have not made any approximations. Let us now expand the
multi-trace correlations according to
(\ref{e-ovNsq-exp-of-corrlns}) and the factorization formula
(\ref{e-factorization}). The SDE at order $1/N^0$ are the
large-$N$ factorized SDE(fSDE) or loop equations(LE). They only
involve the large-$N$ limits of single trace correlations $G_J$
    \beq
    v_i^I S^{J_1 i J_2} G_{J_1 I J_2} = v_i^I \d_I^{I_1 i I_2} G_{I_1}
    G_{I_2} = v^I_i \eta_I^i.
    \label{e-loop-eqns}
    \eeq
Using the notation $L_v = v^I_i L^i_I$ for the vector fields
associated to the infinitesimal changes
(\ref{e-change-of-variable}), these LE may be written $L_v S^J
G_J = v^I_i \eta^i_I$ where $\eta^i_I = \d_I^{I_1 i I_2}
G_{I_1} G_{I_2}$. Here the action of the vector fields on the
moments is $L^i_I G_J = \d_J^{J_1 i J_2} G_{J_1 I J_2}$ and
extends by linearity and the Leibnitz rule to polynomials in
the $G_J$. Moreover, the Lie bracket of two such vector fields
is
    \beq
    [L^{i}_{I}, L^{j}_{J}]=\delta^{J_{1}iJ_{2}}_{J}
        L^{j}_{J_{1}IJ_{2}}-\delta^{I_{1}jI_{2}}_{I} L^{i}_{I_{1}JI_{2}}
    \label{e-lie-alg-of-Ls}
    \eeq
or $[L_u, L_v] = L_w$ where $L_w = w_k^K L_K^k$ and
    \beq
     w^K_k = \sum_{K= K_1 I K_2} ( u^I_i ~
        v^{K_1 i K_2}_k - u^{K_1 i K_2}_k ~ v^I_i).
    \eeq
At ${\cal O}(1/N^2)$, the SDE (one for each $v$ and $n \geq 0$)
involve the $G_J$ as well as the $G^{(2)}_K$'s
    \beq
    v_i^I S^{J_1 i J_2} G^{(2)}_{J_1 I J_2;K_1;\cdots;K_n}
        &=& v_i^I \d_I^{I_1 i I_2} G^{(2)}_{I_1;I_2;K_1;\cdots; K_n}
    \cr &+& v_i^I \sum_{p=1}^n \d_{K_p}^{L_p i M_p} G_{K_1} \cdots
        G_{K_{p-1}} G_{L_p I M_p} G_{K_{p+1}} \cdots G_{K_n}.
    \label{e-subleading-SDE}
    \eeq
We could continue listing the SDE at each order in $1/N^2$, but
we refrain from doing so since they no longer involve the
single trace correlations $G_J$. $G_J$ are the primary objects
of interest in the large-$N$ limit and our goal is to determine
them for a given action $S(A)$. It would be ideal if we could
uniquely determine them by solving the LE (\ref{e-loop-eqns}).
Unfortunately, as was demonstrated in \cite{gsk-loop-eqns} (and
reviewed in \ref{s-underdeterminacy-of-LE}), this is not
possible for many interesting actions $S(A)$, since the LE are
underdetermined. One source of this problem was that there are
vector fields $v$ for which both LHS and RHS of
(\ref{e-loop-eqns}) identically vanish for all\footnote{When we
say `for all $G_J$', we really mean `for all cyclic and
hermitian $G_J$'.} $G_J$, so that the LE for those $v$ are
vacuous. Such vector fields are associated to simultaneous
symmetries of the action and measure in the large-$N$ limit. We
will call such symmetries {\em non-anomalous symmetries} of the
large-$N$ limit\footnote{Anomalous symmetries leave the action
invariant but not the measure.}. Of course, in general, $v$
need not be a symmetry of either action or measure.

We would like to use the ${\cal O}(1/N^2)$ SDE
(\ref{e-subleading-SDE}) to determine the $G_J$ that the LE do
not fix. In principle, (\ref{e-subleading-SDE}) are always
valid. However, (\ref{e-subleading-SDE}) involve the
$G^{(2)}$'s which we do not wish to determine (and most likely
cannot, without also involving the ${\cal O}(\ov{N^4})$ SDE and
so on). Thus, we would like to use the subleading SDE
(\ref{e-subleading-SDE}) only for those $v$ for which the LE
(\ref{e-loop-eqns}) are vacuous. But even these equations would
seem to involve the pesky $G^{(2)}$'s. Fortunately, a
remarkable stroke of good fortune comes to our rescue. Suppose
a vector field $v$ is such that it is a simultaneous symmetry
of the action and measure at $N=\infty$, i.e. $v_i^I S^{J_1 i
J_2} G_{J_1 I J_2} = 0 = v^I_i \eta_I^i$ for all $G_J$. Then we
will show (sections
\ref{s-vol-pres-vfld-ann-terms-in-LE-and-SDE} and
\ref{s-proof-of-ward-id}) that $v$ is also a simultaneous
symmetry at finite $N$, and thence a symmetry at each order in
$1/N^2$:
    \beq
    v_i^I \d_I^{I_1 i I_2} \bra \F_{I_1} \F_{I_2} \F_{K_1} \cdots
    \F_{K_n} \ket =& 0 &= v_i^I S^{J_1 i J_2} \bra \F_{J_1 I J_2} \F_{K_1} \cdots
    \F_{K_n} \ket ~~~~~ \forall ~~~ n=0,1,2,\cdots \cr \cr
    {\rm if~~~~~} v_i^I S^{J_1 i J_2} G_{J_1 I J_2} =& 0 &= v^I_i \eta_I^i
    ~~\forall~~ G_J.
    \eeq
Thus, the terms involving the $G^{(2)}$'s in
(\ref{e-subleading-SDE}) would identically vanish for such $v$
and (\ref{e-subleading-SDE}) would reduce to a set of `Ward'
identities
    \beq
    v_i^I \sum_{p=1}^n \d_{K_p}^{L_p i M_p} G_{K_1} \cdots
        G_{K_{p-1}} G_{L_p I M_p} G_{K_{p+1}} \cdots G_{K_n} = 0~~~~
        \forall ~~~ K_j,~~ n \geq 1.
    \eeq
We call these `Ward' identities (WI) since they are analogues
of the Ward-Takahashi-Slavnov-Taylor identities of Yang-Mills
theory. The latter are a consequence of BRST changes of
variable (\ref{e-brst-transformation}) in functional integrals.
Recall that the BRST transformations are also non-anomalous in
the sense that they leave both the gauge fixed Yang-Mills
action and measure invariant.

These Ward identities can be written more compactly as
    \beq
    L_v (G_{K_1} \cdots G_{K_n}) =0 ~~~~\forall~~ K_p,~ n \geq 1.
    \eeq
Those for $n > 1$ follow from those for $n=1$ and the Leibnitz
rule. So the WI may be taken as
    \beq
    L_v G_K = 0 ~~\forall~~ K {\rm ~~~provided~~} v
        {\rm ~~is~~ such~~ that~~} v_i^I S^{J_1 i J_2} G_{J_1 I J_2} = 0 = v^I_i
        \eta_I^i ~~~ \forall ~~ G_J.
    \label{e-ward-id}
    \eeq
It is satisfying to see that WI, which arise as a consequence
of non-anomalous symmetries, may be regarded as a special case
of the more general concept of Schwinger-Dyson equations. This
is really a statement about quantum field theory in general,
though we are discussing matrix models here. Traditionally
\cite{itzykson-zuber-qft} Ward-like identities are not regarded
as related to Schwinger-Dyson equations in this manner. As
pointed out in \cite{gsk-loop-eqns} and reviewed in section
\ref{s-underdeterminacy-of-LE}, the factorized large-$N$ SDE
are often insufficient to determine the correlations of a
matrix model. However, when the fSDE are supplemented by the
above WI, it becomes possible to determine many (and possibly
all) the correlations, as we will see in later sections.

%----------------------------------------------
\subsection{The case of a single matrix}
\label{s-single-matrix}
%----------------------------------------------

For a $1$-matrix model with action $\tr S(A) = \tr \sum_{1 \leq
n \leq m} S_n A^n$, we use the changes of variable $L_v : A \to
A + \sum_{n \geq -1} v_n A^{n+1}$. A convenient basis is $L_n A
= A^{n+1}, n=-1,0,1,\ldots$. These are familiar from the Lie
algebra of polynomial vector fields on the real line, $L_n =
x^{n+1} \dd{}{x}, n=-1,0,1,\ldots$. Their Lie bracket is
$[L_m,L_n] = (n-m) L_{m+n}$. Equivalently,
    \beq
    L_u = \sum_{n \geq -1} u_n L_n   ~~ {\rm with}
    ~~~ [L_u,L_v]=L_w ~~~ {\rm where~~~} w_k = \sum_{\stackrel{m+n=k}{m,n \geq -1}} (n-m) u_m v_n.
    \eeq
Their action on the moments $G_n = \lim_{n \to \infty} \bra
\F_n \ket$ is $L_k G_n = n G_{k+n}$. Here $\F_n = \Ntr A^n$.
The moments are real by hermiticity of $A$. The LE are
    \beq
    \sum_{k \geq -1} v_k \sum_{n=0}^m n S_n G_{k+n}
    ~=~ \sum_{k \geq -1}v_k \eta_k = \sum_{k \geq -1} v_k \sum_{\stackrel{p+q =k}{p,q \geq
    0}}  G_p G_q.
    \label{e-1-mat-loop-eq}
    \eeq
The LHS is the expectation value of change in action $\sum_{k
\geq -1} v_k L_k \sum_{1 \leq n \leq m} S_n G_n$ while the RHS
is the expectation value of the infinitesimal change in measure
in the large-$N$ limit.

The only vector fields for which the $N = \infty$ expectation
value of the change in measure vanishes, are translations $A
\to A + v_{-1} {\bf 1}$. To see this note that the change in
measure term is
    \beq
    \sum_{k \geq -1} v_k \sum_{\stackrel{p+q=k}{p,q \geq 0}}
    G_p G_q = v_0 + 2 v_1 G_1 + v_2(2 G_2 + G_1^2) + \cdots
    \eeq
If this is to vanish for arbitrary\footnote{The only
constraints on the moments are that they be real and satisfy
the moment inequalities, i.e. that the Hankel matrix $g_{i,j} =
G_{i+j}$ be a positive matrix.} real $G_1, G_2, \cdots$, then
we must have $v_0=v_1=v_2= \cdots =0$. Thus only $v_{-1}$ can
be non-vanishing, which corresponds to a translation.

The only action for which translations are a symmetry in the
large-$N$ limit is the trivial action, $S(A) =$ constant: the
expectation value of change in action under a translation is
    \beq
    \sum_{k \geq -1} v_k \sum_{n=0}^m n S_n G_{n+k} = v_{-1}
    \sum_{n=0}^m n S_n G_{k+n} = v_{-1} (S_1 + 2 S_2 G_1 + 3 S_3 G_2
        + \cdots + m S_m G_{m-1}).
    \eeq
If this must vanish $\forall$ $G_n$, we must have $S_1 = S_2 =
\cdots = S_m =0$, i.e. a trivial action. Thus, for a $1$-matrix
model, we have no infinitesimal simultaneous symmetries of
action and measure in the large-$N$ limit. Consequently, there
are no WI to supplement the LE with.

This leaves a small mystery for $1$-matrix models. As discussed
in \cite{gsk-loop-eqns}, the LE (\ref{e-1-mat-loop-eq}) of a
$1$-matrix model are underdetermined. They do not fix $G_1,
G_2, \cdots G_{m-2}$. The higher moments are fixed in terms of
these by the LE. How are the first few moments to be determined
if there are no WI to supplement the LE? Of course, for a
$1$-matrix model, there are alternative techniques such as
solving the integral equation for the eigenvalue density
\cite{bipz}. For multi-matrix models, the LE are often more
severely underdetermined (there are an infinite number of
moments that are not fixed). Remarkably, for multi-matrix
models, where no alternative systematic method of solution
exists, the WI {\em do} alleviate the underdeterminacy of the
LE (section \ref{s-ward-id}).

%-----------------------------------------------------------
\subsection{Underdeterminacy of multi-matrix loop equations}
\label{s-underdeterminacy-of-LE}
%-----------------------------------------------------------

The multi-matrix LE (\ref{e-loop-eqns}), can also be written as
$S^{J_1 i J_2} G_{J_1 I J_2} = \d_I^{I_1 i I_2} G_{I_1}
G_{I_2}$ for each $i$ and $I$. This form, where we take the
monomial basis $L^i_I$ for vector fields $L_v$ is convenient
for our current discussion. In general, these LE are
underdetermined, as found in section 2.2 of
Ref.\cite{gsk-loop-eqns}. Part of the reason for this
underdeterminacy is the presence of non-anomalous symmetries of
action and measure. First, we establish that the LE for given
$I,i$ can be regarded as a system of inhomogeneous linear
equations for higher rank correlations with lower rank ones
possibly appearing non-linearly. From the LE, it is clear that
if there are any correlations appearing on the LHS, they will
be of a higher rank than the ones on the RHS. More precisely,
suppose the action is an $m^{\rm th}$ order polynomial (i.e.
there is a non-vanishing coupling tensor $S^K$ with $|K|=m$).
LHS of the LE for given $I,i$ (if it is
non-trivial\footnote{Even if there is an $S^K \ne 0$ with
$|K|=m$, it may still happen that for some choice of $I$ and
$i$, the coefficients of all correlations of rank $|I|+m-1$ on
the LHS of the LE vanish. An example is the Gaussian+YM
2-matrix model LE with $m=4$ and empty $I$, given later in this
section. Thus, it is not true in general that the maximal rank
correlation appearing in a LE has rank $|I|+m-1$. This
possibility, which is special to multi-matrix models and has no
analogue for 1-matrix models was overlooked in Ref.
\cite{gsk-loop-eqns}.}), involves correlations only linearly
and with a rank between $|I|$ and $|I|+m-1$, while the highest
rank correlation on the RHS has rank $|I|-1$ . Even if the LHS
vanishes, the highest rank correlation in the LE still appears
linearly, but now on the RHS, and has rank $|I|-1$. However, in
many cases, we find that this system of linear equations is
inadequate to determine all $G_I$.

Let us illustrate this with a Gaussian + Yang-Mills $2$-matrix
model $\tr S(A) = \tr [\frac{m}{2} (A^2_1 + A^2_2)- \ov{2\a}
[A_1,A_2]^2] $. The matrix integrals for this model converge,
and the correlations make rigorous sense and could for instance
be measured numerically. So it makes sense to try to find them
by solving the LE. In this case, the cyclically symmetric
coupling tensors are
    \beq
    S^{11} = S^{22}= {m \over 2},  S^{1122} = S^{2112} = S^{2211} = S^{1221} = 1/{(4\a)} {\rm ~~and~~}
    S^{1212} = S^{2121} = -1/{(2\a)}.
    \eeq
The LE are
    \beq
    i=1: && mG_{I1} - \a^{-1} (2G_{I212}-G_{I221}-G_{I122})
        = \d^{I_1 1 I_2}_I G_{I_1} G_{I_2} {\rm ~~and} \cr
    i=2: && mG_{I2}- \a^{-1}(2G_{I121}-G_{I112}-G_{I211})
        = \d^{I_1 2 I_2}_I G_{I_1}G_{I_2}.
    \eeq
For $I = \emptyset$, the LE say\footnote{This example
illustrates that even in a quartic model (m=4), the LE may
determine correlations of rank $m-2=2$ or less.} $G_1 = G_2 =
0$. The LE for $|I| =1$ relate $2$- and $4$-point correlations:
    \beq
    I=1: &&
    mG_{11}- \ov{\a} (2G_{1212}-2G_{1221}) = 1 ~\&~
    mG_{12}- \ov{\a} (2G_{1121}-G_{1112}-G_{1211}) = 0 \cr
    I=2: &&
    mG_{21}- \ov{\a} (2G_{2212}-G_{2221}-G_{2122}) = 0 ~\&~
    mG_{22}- \ov{\a} (2G_{1212}-2G_{1221}) = 1.
    \eeq
They give the conditions $G_{12} = G_{21} =0, G_{11} = G_{22}$
and $m G_{11} = 1 + \ov{\a} (2 G_{1212} - 2 G_{1122})$. They do
not determine $G_{11}$ and give only one relation among the $6$
independent rank-4 moments. LE with $|I| =2$ relate $3$- and
$5$-point correlations (since we already found $G_i =0$ .)
    \beq
    I=11: &&    mG_{111}- \ov{\a} (2G_{11212}-2G_{11122}) = 0 ~~~\&~~~
         mG_{112} = 0. \cr
    I=12: &&
        mG_{121}-\ov{\a}(G_{12122}-G_{11222}) = 0 ~~~\&~~~
         mG_{122}- \ov{\a} (G_{11212}-G_{11122})= 0 \cr
    I=21: &&
        mG_{211}- \ov{\a} (G_{12122}-G_{11222}) = 0 ~~~\&~~~
        mG_{212}- \ov{\a} (G_{11212}-G_{11122}) = 0 \cr
    I=22: &&
        mG_{221}  = 0 ~~~\&~~~
        mG_{222}- \ov{\a} (2G_{12122}-2G_{11222}) = 0.
    \eeq
The $|I|=2$ LE imply that all $G_{ijk}$ vanish and give two
relations among the $8$ independent\footnote{$c(n,\La)$ denotes
the dimension of the space of cyclically symmetric hermitian
tensors of rank $n$ in a $\La$ matrix model. A formula for
$c(n,\La)$ is given in appendix \ref{a-cyclic-tensors-rank-n}.}
5th rank moments, $G_{12122} = G_{11222}$ and $G_{11212} =
G_{11122}$. The $|I|=3$ LE relate $2$- $4$- and $6$-point
moments (we omit those equations that contain no new
information)
    \beq
    I=111: &&
        mG_{1111}- \ov{\a} (2G_{111212}- 2G_{111122}) = 2G_{11}
        ~~~ {\rm and}~~~  mG_{1112} = 0  \cr
    I=112: && 2G_{112212} = G_{111222} + G_{112122} \cr
        && mG_{1122}- \ov{\a} (2G_{111212}-G_{112112}-G_{111122}) = G_{11} \cr
    I=121: &&
        2G_{121212} = G_{112122} + G_{112212} \cr
        && mG_{1212} =  {2 \over \a} (G_{112112} -  G_{111212}) \cr
    I=211: &&
        2G_{112122} = G_{112212} + G_{111222} \cr
%        && mG_{1122}- \ov{\a} (2G_{111212}-G_{111122}-G_{112112}) = G_{11} \cr
    I=222: &&
        mG_{1222} = 0 {\rm ~~~ and ~~~}
        mG_{2222}- \ov{\a} (2G_{121222}-2G_{112222}) = 2G_{22} \cr
    I=122: &&
        mG_{1122}- \ov{\a} (2G_{121222}-G_{112222}-G_{122122}) = G_{22} \cr
%        && 2G_{112212}= G_{112122} + G_{111222} \cr
    I=212: &&
        mG_{1212} =  {2 \over \a} (G_{122122} - G_{121222}).
%        && 2G_{121212} = G_{112212} + G_{112122} \cr
%    I=221: &&
%        mG_{1122}- \ov{\a} (2G_{121222}-G_{122122}-G_{112222}) = G_{22} \cr
%        && 2G_{112122} = G_{111222} + G_{112212}
    \eeq
However, these $11$ equations (even if all are independent),
are inadequate to find the $c(n=6,\La=2) = 14$ independent
rank-$6$ correlations, let alone the unknown $2$ \& $4$-point
correlations.

Similarly, consider a Yang-Mills $2$-matrix model $\tr S(A) =
-\ov{2\a} \tr [A_1,A_2]^2$. The matrix integrals do not
converge here due to the flat directions in the commutator
squared action. So our derivation of the LE and WI are not
strictly valid in the case, though they can be considered
formally. In particular, it is not clear that the LE and WI
form a consistent system of equations for this model. Nor is it
clear how one could check an answer for a particular
correlation, say by Monte Carlo integration, since the matrix
integrals do not converge. Nevertheless, we can consider the LE
formally here in order to show that they are underdetermined.
The LE are
    \beq
    \ov{\a} v_1^I (G_{I122} + G_{I221} - G_{I212})
    + \ov{\a} v_2^I (G_{I211} + G_{I112} - G_{I121}) = [v_1^I \d_I^{I_1 1 I_2}
    + v_2^I \d_I^{I_1 2 I_2}] G_{I_1} G_{I_2}.
    \eeq
Since $v_i^I$ are arbitrary, we get a pair of LE (for each word
$I$ with $|I| \geq 0$)
    \beq
    G_{I221} + G_{I122} -2G_{I212} =
    \alpha\delta^{I_{1}1I_{2}}_{I}G_{I_{1}}G_{I_{2}}, ~~
    G_{I112} + G_{I211} - 2G_{I121} = \a \d^{I_{1}2I_{2}}_{I}G_{I_{1}}G_{I_{2}}.
    \eeq
All correlations of rank $1$ or $2$ are undetermined. In
addition, taking $I = \emptyset$ does not give any relation for
third rank moments, since the LHS of the LE identically vanish
on account of cyclic symmetry. As for rank-$4$ moments, we get
only one relation $2 G_{1212} - 2 G_{1122} = -\a$, from the LE,
which is inadequate to fix the $6$ independent $4^{\rm th}$
rank correlations.

Similarly, the LE of the Chern-Simons $3$-matrix model $\tr
S(A) = {2i \ka \over 3} \eps^{ijk} \tr A_i A_j A_k$ and Mehta
2-matrix model \cite{Mehta:1981xt} $\tr[c A_1 A_2 + (g/4)
(A_1^4 + A_2^4)]$ are underdetermined (\S 2.2 of
\cite{gsk-loop-eqns}).

%----------------------------------------------------------
\section{Measure preserving transformations}
\label{s-meas-pres-transformations}
%----------------------------------------------------------

Our aim in this section is to determine the vector fields $L_v:
A_i \to A_i + v_i^I A_I$ under whose action the matrix model
measure is invariant. We call such transformations measure or
volume preserving. These vector fields are universal in the
sense that they are independent of the choice of action $S(A)$.
They can only depend on the size of the matrices ($N$), the
number of matrices ($\La$) and on the ensemble from which the
matrices are drawn (hermitian in our case).

The main result of this section\footnote{Section
\ref{s-charac-mean-pres-tr}, where this is established is a bit
long and can be skipped in a first reading.} is that vector
fields $L_v = v^I_i L^i_I$ satisfying
(\ref{e-meas-pres-vect-fld}) are measure preserving for any
$N$. In the large-$N$ limit these are the only ones, but for
finite $N$ there could be more. In particular, a symmetry of
the matrix model measure at $N=\infty$ is automatically a
symmetry of the measure at finite $N$ and consequently at each
order in $1/N^2$. A simpler sufficient condition for a vector
field to be measure preserving is
    \beq
    v^{(I)i (J)}_i + v^{(J) i (I)}_i = 0, ~~~~ \forall ~~ I,J.
    \eeq
Here $(\cdots)$ denotes cyclic symmetrization
(\ref{e-cyclic-symmetrization}). Measure preserving vector
fields form an infinite dimensional Lie algebra for $\La > 1$
with Lie bracket (\ref{e-lie-alg-of-Ls}) (see section
\ref{s-vol-pres-vfld-form-lie-alg}). For $\La=1$, it is a
$1$-dimensional abelian Lie algebra consisting of translations
$A \to A + v_{-1} {\bf 1}$ (see section \ref{s-single-matrix}).

%-----------------------------------------
\subsection{Change in measure due to action of (homogeneous) vector fields}
\label{s-change-in-meas-homogeneous-vfld}
%-----------------------------------------

If all $I$ appearing in the components $v_i^I$ of the vector
field $L_v = v^I_i L_I^i$ have the same length $|I|$, then we
will call such a vector field homogenous of rank $|I|$. The
variation of the measure under (a not necessarily homogenous)
infinitesimal change of variable,
    \beq
    A_{i}\rightarrow A'_{i} = A_{i} + v_{i}^{I} A_{I}
    \eeq
is the first order term in the expansion of the determinant of
the Jacobian $J$ in powers of $v_{i}^{I}$
    \beq
        J = \det \bigg[ \frac{\pdr A_{ic}^{\pr d}} {\pdr A_{ja}^b}
        \bigg]
        =\det [\d^i_j \d^a_c \d^d_b + v_i^I
        \d_I^{I_1 iI_2} \Phi_{I_1} \Phi_{I_2} + {\cal O}(v^2)]
    = 1 + N^2 v_i^I \d_I^{I_1 iI_2} \Phi_{I_1} \Phi_{I_2} + O(v^2).
    \eeq
Here $\Phi_I = \Ntr A_I$. Thus infinitesimally, the change in
the measure per $N^2$ is
    \beq
    {\delta J  \over N^2} = v_{i}^{I}\delta_{I}^{I_{1}iI_{2}}\Phi_{I_{1}}\Phi_{I_{2}}.
    \label{e-inf-change-in-measure}
    \eeq
We want to determine those $v_{i}^{I}$'s for which $\d J/N^2
=0$ for all $\Phi_K$ which are cyclically symmetric in $K$ and
hermitian $\Phi^*_K = \Phi_{\bar K}$. So we should set the
coefficients of independent $\F_K$ to zero. Unfortunately, for
finite-$N$, the analysis is complicated by the fact that
$\Phi_K$ are not all independent. Indeed, they are related by
trace identities (analogues of conditions from vanishing of
characteristic polynomial for a single matrix). However, in the
large-$N$ limit, the $\F_I$ (and consequently their expectation
values, $G_I = \bra \F_I \ket$) are independent up to cyclic
symmetry and hermiticity. The expectation value of the
infinitesimal change in measure becomes
    \beq
    \lim_{N \to \infty} \bra {\d J \over N^2} \ket = v_i^I \d_I^{I_1 i I_2} G_{I_1}
    G_{I_2} = v^I_i \eta^i_i \equiv v \cdot \eta.
    \eeq
Setting the coefficient of each independent monomial in the
$G_I$ to zero leads to a characterization of the measure
preserving vector fields $v^I_i$ in the large-$N$ limit
(section \ref{s-charac-mean-pres-tr}).

An inspection of $v \cdot \eta$ reveals that it vanishes under
the sum of two homogeneous transformations $\delta A_{i}=
\sum_{|I|=const} v_{i}^{I} A_{I} + \sum_{|J|=const} v_{i}^{J}
A_{J}$ with $|I| \neq |J|$ if and only if it vanishes under
each separately. So without loss of generality, we restrict to
homogenous vector fields.

Notice from (\ref{e-lie-alg-of-Ls}) that the commutator of
homogenous vector fields of rank $p>1$ and $q>1$ is a
homogenous vector field of rank $p+q-1$. Using this, we define
the grading of a homogenous vector field $\sum_{|I| = const}
v_i^I L^{i}_{I}$ as $|I|-1$. With this, the Lie algebra
$\mathcal{L}$ of all $L$'s\footnote{Except translations
$v^\emptyset_i L^i_\emptyset: A_i \to A_i + v_i^\emptyset {\bf
1}$, which are a separate abelian algebra. $\emptyset$ is the
empty word.} becomes a Lie algebra $\mathcal{L} = \bigoplus_{p
\geq 0}\mathcal{L}_{p}$ graded by the non-negative integers,
where
    \beq
    \mathcal{L}_{p} = span \left\lbrace L^{i}_{I}: | I|=p+1 \right\rbrace
    ~~~ {\rm and ~~~}
    \left[\mathcal{L}_{p},\,\mathcal{L}_{q}\right] \subset
    \mathcal{L}_{p+q}.
    \eeq
This can be used to generate homogeneous higher rank volume
preserving vector fields from ones with lower rank, provided
the latter do not form an abelian Lie algebra.

%------------------------------------------------
\subsection{Characterization of measure preserving vector fields for $N=\infty$}
\label{s-charac-mean-pres-tr}
%------------------------------------------------

Roughly, the condition that a vector field be volume preserving
becomes stronger as $N \to \infty$, since the number of
independent trace invariants grows in this limit. So there are
potentially a lot more volume preserving vector fields at
finite-$N$ than at $N=\infty$. Fortunately, the condition that
a vector field be volume preserving at $N=\infty$ will be seen
to be a sufficient condition for it to be volume preserving at
any finite $N$. In this manner, we will establish that if the
change of measure term in the LE (\ref{e-loop-eqns}) vanishes
for a given vector field $v$, then it also
vanishes\footnote{There likely exist vector fields preserving
the measure at finite-$N$ but not at $N=\infty$.} in the
finite-$N$ Schwinger-Dyson equations (\ref{e-SDE}) and indeed
at each order in $1/N^2$ for the same vector field $v$. To
characterize volume preserving vector fields in the large-$N$
limit we must solve the equations $v \cdot \eta =0$ for
$v^I_i$. Let us begin with homogeneous vector fields of lowest
rank.

\pt {\bf Constant shift:} $\d A_i = v_i {\bf 1}$. In this case
$v \cdot \eta =0$. So all homogeneous vector fields of rank
zero are symmetries of the measure. This reflects translation
invariance of the measure.

\pt {\bf Linear transformation:} $\delta A_{i}=v^{j}_{i}A_{j}$
are measure preserving iff they are traceless:
    \beq
    v \cdot \eta ~=~  v^{j}_{i}\delta^{i}_{j}=v^{i}_{i} {\rm ~~~so~~~}
    v \cdot \eta =0 ~\Leftrightarrow~ v {\rm ~~is~ traceless~~} v^{i}_{i}=0.
    \eeq

\pt {\bf Quadratic:} $\delta A_{i}=v^{jk}_{i}A_{j}A_{k}$. In this case,
    \beq
    v \cdot \eta &=& v^{jk}_{i}(\delta^i_k G_j + \delta^{i}_{j} G_{k})
        = v^{ji}_{i} G_{j} + v^{ik}_i G_k =
        (v^{ij}_{i}+v^{ji}_{i}) G_j
    \eeq
Thus quadratic vector fields that preserve the measure must
satisfy $v^{ij}_{i}+v^{ji}_{i}=0$.

\pt {\bf Cubic:} $\delta A_{i}=v^{jkl}_i A_j A_k A_l$
    \beq
        v \cdot \eta = v^{jkl}_{i}(\delta^{imn}_{jkl} G_{mn}
        + \delta^{min}_{jkl} G_{m} G_{n}+\delta^{mni}_{jkl} G_{mn})
    = (v^{imn}_{i} + v^{mni}_{i}) G_{mn} + v^{min}_{i} G_{m} G_{n}.
    \eeq
This must vanish for all cyclic and hermitian $G_K$. The linear
term in $G$'s is cyclically symmetric in $mn$, so it is not
necessary that the coefficient of $G_{mn}$ and $G_{nm}$
separately vanish. Rather, only the cyclic projection of
$G_{mn}$'s coefficient must vanish. Similarly, the quadratic
term in $G$'s is symmetric under $m \leftrightarrow n$ so only
the symmetric projection of its coefficient must vanish.
Moreover, hermiticity implies $G_j^* = G_j$ and $G_{mn}^* =
G_{nm} = G_{mn}$, so all $1$- and $2$-point correlations are
real. We need not worry about setting the coefficients of their
imaginary parts to zero. Thus $v \cdot \eta$ vanishes
identically if and only if
    \beq
    v^{imn}_i + v^{inm}_i + v^{mni}_i +v^{nmi}_i =0
        { \rm ~~~ and ~~~} v^{min}_i +v^{nim}_i =0.
    \eeq
We can write this more succinctly as
    \beq
    v^{i(mn)}_i + v^{(mn)i}_i=0 { \rm ~~~ and ~~~} v^{min}_i + v^{nim}_i =
    0.
    \eeq
Here we introduced the cyclic symmetrization operation
$(\cdots)$ which is defined as\footnote{$C_{|J|}$ is the cyclic
group of order $|J|$. Note that we do not divide by the number
of terms.}
    \beq
    v_i^{(J) i K} = \sum_{\pi \in C_{|J|}} v_i^{\pi(J) i K}, ~~~
    v_i^{J i (K)} = \sum_{\sigma \in C_{|K|}} v_i^{J i \sigma(K)},
    ~~~
    v_i^{(J) i (K)} = \sum_{\stackrel{\pi \in C_{|J|}}{\sigma \in C_{|K|}}} v_i^{\pi(J) i
    \sigma(K)}.
    \label{e-cyclic-symmetrization}
    \eeq

\pt {\bf Quartic:} $\delta A_{i}=v^{jklm}_{i}A_{j}A_{k}A_{l}A_{m}$
    \beq
    v \cdot \eta &=& v^{jklm}_{i}(\delta^{ipqr}_{jklm} G_{pqr} +
    \delta^{piqr}_{jklm} G_{p} G_{qr}+\delta^{pqir}_{jklm} G_{pq} G_{r}
    + \delta^{pqri}_{jklm} G_{pqr})
    \cr
    &=&(v^{ipqr}_{i}+v^{pqri}_{i}) G_{pqr}+v^{piqr}_{i} G_{p} G_{qr}
        + v^{pqir}_{i} G_{pq} G_{r} \cr
%    &=&(v^{ipqr}_{i} + v^{pqri}_i) G_{pqr} + v^{piqr}_{i} G_{p} G_{qr}
%        + v^{qrip}_{i} G_{qr} G_{p}     \cr
    &=&(v^{ipqr}_{i}+v^{pqri}_{i}) G_{pqr}+(v^{piqr}_{i}+v^{qrip}_{i}) G_{p} G_{qr}.
    \eeq
Here, the quadratic term is cyclically symmetric in $qr$, so
its coefficient must be cyclically symmetrized in $qr$.
Similarly, the linear term is cyclically symmetric in $pqr$ so
we must cyclically symmetrize its coefficient in $pqr$.
However, there is a further subtlety that we must address:
$G_K$ are complex numbers, but their real and imaginary parts
are related to those of $G_{\bar K}$ via $G_K^* = G_{\bar K}$.
Hermiticity implies that $G_p$ and $G_{qr}$ are real, so for
the quadratic term, it is necessary and sufficient that
($v_i^I$ are real)
    \beq
    v^{pi (qr)}_i + v^{(qr)ip}_i = 0.
    \eeq
On the other hand, $G_{pqr}^* = G_{rqp}$. So $\Re G_{pqr} = \Re
G_{rqp}$ and $\Im G_{pqr} = - \Im G_{rqp}$. So it is necessary
and sufficient to set the coefficients of $\Re G_{pqr} + \Re
G_{rqp}$ and $\Im G_{pqr} - \Im G_{rqp}$ to zero separately:
    \beq
    v^{i (pqr)}_i + v^{(pqr)i}_i + v^{i (rqp)}_i + v^{(rqp)i}_i =0
    {\rm ~~ and ~~}
    v^{i (pqr)}_i + v^{(pqr)i}_i - v^{i (rqp)}_i - v^{(rqp)i}_i =0
    ~~~ \forall ~~ pqr.
    \eeq
However, these two conditions are equivalent (by adding and
subtracting) to the single condition
    \beq
    v^{i (pqr)}_i + v^{(pqr)i}_i = 0 ~~~ \forall ~~~ pqr.
    \eeq
Therefore $v \cdot \eta$ vanishes if and only if
    \beq
    v^{i (pqr)}_i + v^{(pqr)i}_i=0 ~~~
    {\rm and} ~~~ v^{pi (qr)}_i + v^{(qr)ip}_i = 0.
    \eeq

\pt {\bf Quintic:} For a rank $5$ vector field $A_i \to A_i +
v_i^{pqrst} A_{pqrst}$ to be volume preserving we need
    \beq
    v \cdot \eta = \sum_{pqrs} (v_i^{ipqrs} + v_i^{pqrsi}) G_{pqrs}
    + (v_i^{piqrs} + v_i^{qrsip}) G_p G_{qrs} + (v_i^{pqirs} + v_i^{rsipq}) G_{pq}
    G_{rs} = 0.
    \eeq
Since moments of different ranks are independent, $v \cdot \eta
= 0$ iff the following three equations are satisfied (we have
cyclically symmetrized as in previous cases in order to reduce
to a sum over equivalence classes under cyclic symmetry, which
is denoted $\sim$)
    \beq
    \sum_{pqrs/\sim} (v_i^{i(pqrs)} + v_i^{(pqrs)i}) G_{pqrs} = 0,
    &&
    \sum_{p,qrs/\sim} (v_i^{pi(qrs)} + v_i^{(qrs)ip}) G_p G_{qrs} = 0, \cr
    {\rm and} &&
    \sum_{pq/\sim, rs/\sim} (v_i^{(pq)i(rs)} + v_i^{(rs)i(pq)}) G_{pq} G_{rs} = 0.
    \label{e-quintic-conditions}
    \eeq
It remains to take care of the relations imposed by hermiticity
to select the independent monomials. Consider the first
equation in (\ref{e-quintic-conditions}). By hermiticity $\Re
G_{pqrs} = \Re G_{srpq}$ and $\Im G_{pqrs} = - \Im G_{srqp}$.
So we further restrict the sum to equivalence classes under
reversal of order of indices. We will denote the combination of
the quotient by cyclic symmetrization and reversal of order of
indices by the symbol $\sim^\pr$. Then the first condition in
(\ref{e-quintic-conditions})  becomes the pair
    \beq
    \sum_{pqrs/\sim^\pr} \bigg[ v_i^{i(pqrs)} + v_i^{(pqrs)i} + v_i^{i(\overline{pqrs})}
        + v_i^{(\overline{pqrs})i}\bigg] \Re G_{pqrs} &=& 0  {\rm ~~~and} \cr
    \sum_{pqrs/\sim^\pr} \bigg[ v_i^{i(pqrs)} + v_i^{(pqrs)i} - v_i^{i(\overline{pqrs})}
        - v_i^{(\overline{pqrs})i}\bigg] \Im G_{pqrs} &=& 0.
    \eeq
Since the sum is over independent moments, we set the
coefficients to zero and get
    \beq
    v_i^{i(pqrs)} + v_i^{(pqrs)i} + v_i^{i (\overline{pqrs})} + v_i^{(\overline{pqrs})i} &=& 0
    {\rm ~~~~ and }\cr
    v_i^{i(pqrs)} + v_i^{(pqrs)i} - v_i^{i(\overline{pqrs})} - v_i^{(\overline{pqrs})i} &=&
    0,
    \eeq
for each equivalence class $pqrs$ under the relation
$\sim^\pr$. However, this pair is equivalent to
    \beq
    v_i^{i(pqrs)} + v_i^{(pqrs)i} = 0 ~~~ \forall ~~~{\rm cyclic ~~equivalence
    ~~classes~~}
    pqrs/\sim.
    \eeq
A similar analysis of the last two conditions in
(\ref{e-quintic-conditions}) using hermiticity ($G_p^* = G_p$
and $\Re G_{qrs} = \Re G_{\overline{qrs}}$ and $\Im G_{qrs} = -
\Im G_{\overline{qrs}}$ and $G_{pq}^* = G_{pq}$) allows us to
identify the coefficients of independent moments and set them
to zero. When the dust settles, the necessary and sufficient
conditions for a $5^{\rm th}$ rank tensor to be measure
preserving are
    \beq
    v_i^{i (pqrs)} + v_i^{(pqrs) i} =0; ~~~
    v_i^{pi(qrs)} + v_i^{(qrs) ip} =0; ~~~
    v_i^{(pq)i(rs)} + v_i^{(rs)i(pq)} =0 ~~~ \forall~ p,q,r,s.
    \eeq
We see that so far, the hermiticity relations between the
$G_K$, though taken into account, did not make their presence
felt in the final answer. This simplification is due to $G_i$
and $G_{ij}$ being real. The hermiticity relations will play a
role in the necessary and sufficient conditions for rank $7$
and higher vector fields to be measure preserving. This is
because it is the first case where $\eta^i_I$ involves
quadratic monomials in moments where both factors can be
complex, e.g. $G_{pqr} G_{stu}$. This leads to complications
which we now deal with in the general case.

\pt {\bf Rank $n$:} In the general case, $\delta
A_{i}=v^{j_{1}\ldots j_{n}}_{i}A_{j_{1}}\ldots A_{j_{n}}$ and
$v \cdot \eta = v_i^{I_1 i I_2} G_{I_1} G_{I_2}$ is a quadratic
polynomial in moments. The necessary and sufficient conditions
on $v$ for $v \cdot \eta=0$ are got by selecting the
independent monomials and setting their coefficients to zero.
We use three relations: $(a)$ commutativity of products $G_I
G_J = G_J G_I$, $(b)$ cyclicity $G_I = G_J$ if $I,J$ are
cyclically related and $(c)$ hermiticity $G_I = G_{\bar I}^*$
or $\Re G_I = \Re G_{\bar I}$ and $\Im G_I = - \Im G_{\bar I}$.
$(a) \implies$ we must symmetrize the coefficients in $I_1$ and
$I_2$ and restrict the sum over $I_1$ and $I_2$ to include only
(say) $G_{I_1} G_{I_2}$ and not $G_{I_2} G_{I_1}$ (this is
denoted $\sum^\pr$).
    \beq
    v \cdot \eta = 0 ~~\Leftrightarrow~~
    {\sum_{I_1, I_2}}^\pr (v_i^{I_1 i I_2} + v_i^{I_2 i I_1}) G_{I_1} G_{I_2} =0.
    \eeq
Relation $(b)$ means we must cyclically symmetrize coefficients
in $I_1$ and $I_2$ and further restrict the sum to cyclic
equivalence classes of $I_1$ and $I_2$ (denoted $I_1/\sim$)
    \beq
    v \cdot \eta = 0 ~~\Leftrightarrow~~
    {\sum_{\stackrel{I_1/\sim}{I_2/\sim}}}^\pr \bigg[v_i^{(I_1) i (I_2)}
        + v_i^{(I_2) i (I_1)} \bigg] G_{I_1} G_{I_2} =0.
    \eeq
Implementing $(c)$ is more tricky. We must identify monomials
that are independent after accounting for hermiticity. Taking
$\Re$ \& $\Im$ parts($v_i^I \in {\bf R}$), we write $v \cdot
\eta =0$ as the pair
    \beq
    {\sum_{\stackrel{I/\sim}{J/\sim}}}^\pr \bigg[v_i^{(I) i (J)}
        + v_i^{(J) i (I)} \bigg] (\Re G_I \Re G_J - \Im G_I \Im
        G_J) &=& 0 {\rm ~~~and}\cr
    {\sum_{\stackrel{I/\sim}{J/\sim}}}^\pr \bigg[v_i^{(I) i (J)}
        + v_i^{(J) i (I)} \bigg] (\Re G_I \Im G_J + \Re G_J \Im
        G_I) &=& 0.
    \eeq
The last two terms can be combined. Hermiticity $\implies$ $\Re
G_I = \Re G_{\bar I}, \Im G_I= -\Im G_{\bar I}$. So $\Re G_I
\Re G_J$ is independent of $\Im G_I \Im G_J$ and we can set
each part to zero separately. Thus $v \cdot \eta = 0$ iff
    \beq
    {\sum_{\stackrel{I/\sim}{J/\sim}}}^\pr \bigg[v_i^{(I) i (J)}
        + v_i^{(J) i (I)} \bigg] \Re G_I \Re G_J = 0 &,&
    {\sum_{\stackrel{I/\sim}{J/\sim}}}^\pr \bigg[v_i^{(I) i (J)}
        + v_i^{(J) i (I)} \bigg] \Im G_I \Im G_J = 0, \cr {\rm and~~~~}
    {\sum_{\stackrel{I/\sim}{J/\sim}}}^\pr \bigg[v_i^{(I) i (J)}
        + v_i^{(J) i (I)} \bigg] \Re G_I \Im G_J &=& 0.
    \eeq
Here the sums include words $I$ as well as their mirror images
$\bar I$, so the monomials such as $\Re G_I \Re G_J$ are not
all independent on account of the hermiticity relations. We
must further restrict the sums to equivalence classes under
reversal of order of letters in a word to get a truly
independent basis for quadratic polynomials. Once this is done
we set the coefficients to zero and find that $v \cdot \eta =
0$ iff (the signs are determined by the hermiticity relations)
    \beq
    v_i^{(I)i(J)} + v_i^{(\bar I)i(J)} + v_i^{(I)i(\bar J)} + v_i^{(\bar I)i(\bar J)}
        + I \leftrightarrow J &=& 0, \cr
    v_i^{(I)i(J)} - v_i^{(\bar I)i(J)} - v_i^{(I)i(\bar J)} + v_i^{(\bar I)i(\bar J)}
        + I \leftrightarrow J &=& 0 {\rm ~~~and }\cr
    v_i^{(I)i(J)} + v_i^{(\bar I)i(J)} - v_i^{(I)i(\bar J)} - v_i^{(\bar I)i(\bar J)}
        + I \leftrightarrow J &=& 0.
    \eeq
These can be slightly simplified to the following three conditions
    \beq
    v_i^{(I)i(J)} + v_i^{(\bar I) i (\bar J)} + I \leftrightarrow J  =
    0,
    v_i^{(\bar I)i(J)} + v_i^{(I) i (\bar J)} + I \leftrightarrow J  =
    0,
    v_i^{(I)i(J)} - v_i^{(I) i (\bar J)} + I \leftrightarrow J  = 0.
    \label{e-meas-pres-vect-fld}
    \eeq
Thus, a homogeneous vector field $v_i^I$ of rank $n$ is volume
preserving at $N=\infty$ ($v \cdot \eta =0$), iff conditions
(\ref{e-meas-pres-vect-fld}) are satisfied for each multi-index
$I$ and $J$ such that $|I|+|J|=n-1$. Since conditions
(\ref{e-meas-pres-vect-fld}) are somewhat lengthy (though easy
to remember), it is pertinent to add that a sufficient (but in
general not necessary) condition for $v$ to be measure
preserving is
    \beq
    v^{(I)i (J)}_i + v^{(J) i (I)}_i = 0 ~~~ \forall~~ I,J {\rm ~~with~~}
    |I|+|J|=n-1.
    \label{e-meas-pres-vect-fld-suff-cond}
    \eeq
More explicitly, this sufficient condition may be written as
($[n]$ is the greatest integer part of $n$)
    \beq
    v^{i (j_{1}\ldots j_{n-1})}_i
        + v^{(j_{1}\ldots j_{n-1}) i}_i &=& 0, \cr
    v^{j_{1}i (j_{2}\ldots j_{n-1})}_i
        + v^{(j_{2}\ldots j_{n-1})ij_{1}}_i &=& 0, \cr
    v^{(j_{1}j_{2})i (j_{3}\ldots j_{n-1})}_i +
        v^{(j_{3}\ldots j_{n-1}) i (j_{1}j_{2})}_i &=& 0, \cr
    &\vdots& \cr
        v_i^{(j_1 j_2 \cdots j_{[{n-1 \over 2}]}) i (j_{[{n+1 \over 2}]} \cdots j_{n-1})}
        + v_i^{(j_{[{n+1 \over 2}]} \cdots j_{n-1}) i (j_1 j_2 \cdots j_{[{n-1 \over 2}]}) } &=&
        0.
    \eeq
In fact, (\ref{e-meas-pres-vect-fld-suff-cond}) is both
necessary and sufficient for vector fields of rank $\leq 6$.

Now, it is easy to see that a volume preserving vector field
for $N=\infty$ is automatically volume preserving for finite
$N$. Suppose $v^I_i$ is such that $v \cdot \eta = v^I_i
\d_I^{I_1 i I_2} G_{I_1} G_{I_2} =0$, or equivalently, such
that conditions (\ref{e-meas-pres-vect-fld}) are satisfied.
Then it will automatically satisfy ${\d J \over N^2} = v^I_i
\d_I^{I_1 i I_2} \F_{I_1} \F_{I_2} =0$. For, all we needed for
$v \cdot \eta=0$ was commutativity of the product of two
$G_I$'s, cyclic symmetry and hermiticity of the $G_I$. All
these properties are also satisfied by the $\F_I$'s. Of course,
there are likely to be vector fields other than those
satisfying (\ref{e-meas-pres-vect-fld}) (i.e. $v \cdot \eta \ne
0$) for which ${\d J \over N^2} =0$.

%-----------------------------------------
\subsection{Volume preserving vector fields annihilate measure terms in LE and
SDE}
\label{s-vol-pres-vfld-ann-terms-in-LE-and-SDE}
%-----------------------------------------

Thus far, we have shown that vector fields characterized in
(\ref{e-meas-pres-vect-fld}) leave the measure invariant $\d
J/N^2 = v_i^I \d_I^{I_1 i I_2} \F_{I_1} \F_{I_2} =0$ both for
finite and infinite $N$. Moreover, they were all the vector
fields that left the measure invariant for $N=\infty$: $v_i^I
\d_I^{I_1 i I_2} G_{I_1} G_{I_2} = v_i^I \eta^i_I = 0$. In
other words, the RHS of the LE (\ref{e-loop-eqns}) identically
vanish iff the vector field $v$ satisfies
(\ref{e-meas-pres-vect-fld}).

On the other hand, multiplying (\ref{e-inf-change-in-measure})
by $\F_{K_1} \cdots \F_{K_n}$ and taking expectation values we
get
    \beq
    v_i^I \d_I^{I_1 i I_2} \bra \F_{I_1} \F_{I_2} \F_{K_1} \cdots \F_{K_n}
    \ket = 0  ~~~ \forall~~~ \F_K ~~{\rm and~~} n =0,1,2,\ldots
    \eeq
provided $v$ satisfy (\ref{e-meas-pres-vect-fld}). Combining
with the result of the previous paragraph, we see that vector
fields for which the RHS of the LE (\ref{e-loop-eqns}) vanish,
also annihilate the change of measure term on the RHS of the
finite-$N$ Schwinger-Dyson equations (\ref{e-SDE}).
Furthermore, multiplying by $N^2$ and letting $N \to \infty$ we
see that the same class of vector fields also annihilate the
change of measure term on the RHS of the ${\cal O}(1/N^2)$ SDE
(\ref{e-subleading-SDE})
    \beq
    v_i^I \d_I^{I_1 i I_2} G^{(2)}_{I_1; I_2; K_1; \cdots ; K_n} =0.
    \eeq
This is a part of the result we needed in section
\ref{s-sde-and-ward-id} to establish the WI (\ref{e-ward-id}).
The other part involves identifying which of these volume
preserving vector fields also leaves the action of a specific
matrix model invariant, a task we will undertake in section
\ref{s-symmetries-of-action}.

%----------------------------------------------
\subsection{Volume preserving vector fields form an infinite dimensional Lie algebra}
\label{s-vol-pres-vfld-form-lie-alg}
%----------------------------------------------

It should be possible, but laborious, to check that the Lie
bracket of two vector fields of the form
(\ref{e-meas-pres-vect-fld}) is again of the same form (we have
checked this for vector fields of some low ranks). But there is
a simpler argument (which uses a much deeper result from
\cite{entropy-var-ppl}) that shows they form a Lie algebra. In
\cite{entropy-var-ppl} it was shown that there is an entropy
function\footnote{However, $\chi$ cannot be expressed as a
formal power series in $G_I$.} $\chi$ such that $L_I^i \chi =
\eta^i_I = \d_I^{I_1 i I_2} G_{I_1} G_{I_2}$. Suppose $L_u,
L_v$ are volume preserving. From our results in sections
\ref{s-charac-mean-pres-tr} and
\ref{s-vol-pres-vfld-ann-terms-in-LE-and-SDE}, this means
$u_i^I \eta^i_I = 0$ and $v_i^I \eta^i_I = 0$. Then we have
$L_u \chi =0$ and $L_v \chi =0$. It follows therefore that
$[L_u,L_v] \chi = (L_u L_v - L_v L_u) \chi =0$. Thus $L_w =
[L_u,L_v]$ is also volume preserving. We conclude that volume
preserving vector fields form a Lie algebra.

{\bf Example:} The measure preserving vector fields
corresponding to linear transformations,
$L_{u}=u^{i}_{j}L^{j}_{i}$ form the $sl_\La({\bf R})$ Lie
algebra for a $\La$-matrix model. We already found (section
\ref{s-charac-mean-pres-tr}) that measure preserving linear
transformations are the traceless ones. Here, we check that
their Lie bracket implied by (\ref{e-lie-alg-of-Ls}) is the
same as the $sl_\La({\bf R})$ Lie algebra.
    \beq
    [L_{u}, L_{v}]&=&u^{i}_{j}v^{k}_{l}[L^{j}_{i},L^{l}_{k}]
    ~=~ u^{i}_{j}v^{k}_{l} \left(\delta^{j}_{k}L^{l}_{i}-\delta^{l}_{i}L^{j}_{k} \right)
    ~=~ u^{i}_{k}v^{k}_{l}L^{l}_{i}-u^{i}_{j}v^{k}_{i}L^{j}_{k}
    ~=~ w^{i}_{l}L^{l}_{i},
    \eeq
where
$w^{i}_{l}=u^{i}_{k}v^{k}_{l}-v^{i}_{k}u^{k}_{l}=\left([u,v]\right)^{i}_{l}$
is just the matrix commutator. Thus $[L_{u}, L_{v}]=L_{[u,v]}$
and the linear symmetries form the Lie algebra $sl_\La({\bf
 R})$.

Moreover, for $\La > 1$ we can show that the space of measure
preserving vector fields is infinite dimensional. It is
sufficient to consider each rank separately. First, the space
of rank-$n$ vector fields $v_i^{i_1 i_2 \cdots i_n}$ is
$\La^{n+1}$ dimensional. For a rank-$n$ vector field to be
measure preserving it is sufficient (though not necessary) that
it satisfy equations (\ref{e-meas-pres-vect-fld-suff-cond}).
There are at most $\La^{n-1}$ such linear equations (if they
were not linearly independent or necessary, there would be even
fewer). Thus, the space of solutions is at least $\La^{n+1} -
\La^{n-1}$ dimensional. This grows exponentially with rank, so
measure preserving vector fields are an infinite dimensional
Lie algebra for $\La > 1$.

%----------------------------------------------------------
\subsection{Explicit examples for $2$ and $3$ matrix models}
\label{s-eg-meas-pres-vfld-2-3-mat-mod}
%----------------------------------------------------------

From sections \ref{s-charac-mean-pres-tr} and
\ref{s-vol-pres-vfld-form-lie-alg}, we know that linear volume
preserving vector fields are traceless matrices $v_i^j$, i.e.
elements of $sl_\La({\bf R})$. This is a $\La^2 - 1$
dimensional space ($3$ dimensional for a $2$-matrix model and
$8$ dimensional for a $3$-matrix model).

A generic quadratic vector field $v_i^{jk}$ in a $\La$-matrix
model is specified by $\La^3$ parameters. But volume preserving
vector fields obey relations given in section
\ref{s-charac-mean-pres-tr}, which restrict the number of
independent coefficients. Let us work out volume preserving
$v_i^{jk}$ for $2$ and $3$-matrix models and determine the
dimension of the space of such vector fields. The condition for
$v_i^{jk}$ to be volume preserving is
$v^{ij}_{i}+v^{ji}_{i}=0$. In a $2$-matrix model this is the
pair of equations
    \beq
    2v^{11}_{1}+v^{21}_{2}+v^{12}_{2} = 0 {\rm ~~~and~~~ }
    2v^{22}_{2}+v^{12}_{1}+v^{21}_{1} = 0.
    \label{e-2-mat-quad-meas-pres-vfld}
    \eeq
So quadratic volume preserving vector fields are the $2^3-2=6$
parameter family
    \beq
    L_v = v^{ij}_k L^k_{ij}
%      v^{11}_2 L^2_{11} + v^{22}_1 L^1_{22}
%      + v^{12}_1 L^1_{12} + v^{21}_1 L^1_{21} + v^{12}_2 L^2_{12}
%      + v^{21}_2 L^2_{21} \cr
%      && - \half \bigg( v^{12}_2 + v^{21}_2 \bigg) L^1_{11}
%      - \half \bigg( v^{12}_1 + v^{21}_1 \bigg) L^2_{22} \cr
    &=&   v^{11}_2 L^2_{11} + v^{22}_1 L^1_{22}
        + v^{12}_1 \bigg[L^1_{12}-\half L^2_{22} \bigg]
        + v^{21}_1 \bigg[L^1_{21}-\half L^2_{22} \bigg] \cr
    &+& v^{12}_2 \bigg[L^2_{12}-\half L^1_{11}
        \bigg] + v^{21}_2 \bigg[ L^2_{21}-\half L^1_{11}
        \bigg].
    \label{e-2-mat-quad-meas-pres-lie-derivatives}
    \eeq
%One can parameterize these vector fields by a $2^3-2=6$
%parameter $v_1^{12}, v_1^{21}, v_2^{12}, v_2^{21}, v_1^{22},
%v_2^{11}$ family with the remaining two given by
%    \beq
%    v^{11}_{1} = -\half (v^{21}_{2}+v^{12}_{2}) {\rm ~~~and ~~~}
%    v^{22}_{2} = -\half (v^{12}_{1}+v^{21}_{1}).
%    \eeq
In a $3$-matrix model there are three independent conditions
    \beq
    2v^{11}_{1}+v^{21}_{2}+v^{31}_{3}+v^{12}_{2}+v^{13}_{3}=0, &&
    2v^{22}_{2}+v^{12}_{1}+v^{32}_{3}+v^{21}_{1}+v^{23}_{3}=0 \cr
    {\rm and~~~} 2v^{33}_{3}+v^{13}_{1}+v^{23}_{2}+v^{31}_{1}+v^{32}_{2}&=&0.
    \eeq
So quadratic volume preserving vector fields are a $3^3-3=24$
parameter family for $\La = 3$.

%----------------------------------------------------------
\section{Transformations that also preserve action}
\label{s-symmetries-of-action}
%----------------------------------------------------------

%----------------------------------------------------------
\subsection{Establishing validity of Ward identities: last step}
\label{s-proof-of-ward-id}
%----------------------------------------------------------

So far, we have identified the vector fields $L_v = v_i^I
L_I^i$ which leave the measure invariant in the large-$N$ limit
and observed that they continue to be measure preserving even
at finite $N$. In order to obtain the WI (\ref{e-ward-id}), we
need to determine which among these $L_v$ are also symmetries
of the action. These are the non-anomalous infinitesimal
symmetries. The answer will, of course, depend on the action of
the matrix model being studied. For the infinitesimal change in
($\N \times$) the action $S(A) = \tr S^J A_J$ to vanish under
$A_i \to A_i + v_i^I A_I$, we need
    \beq
    L_v S^J \F_J = v_i^I S^{J_i i J_2} \F_{J_1 I J_2} = 0.
    \label{e-action-pres-vfld-finite-N}
    \eeq
However, for finite $N$, not all the $\Phi_I$ are independent
even after accounting for cyclicity and hermiticity, due to the
trace identities and other such constraints satisfied by the
$\F_I$. So it is {\em not} straightforward to identify the {\em
necessary} conditions on $v_i^I$. But in the large-$N$ limit we
may treat the $\Phi_I$ as independent variables (up to
cyclicity and hermiticity). Taking expectation values, we must
solve for $v_i^I$ in the equations
    \beq
    v_i^I S^{J_1 i J_2} G_{J_1 I J_2} =0    ~~~ \forall ~~~
        {\rm cyclic ~~and~~ hermitian~~} G_I.
    \label{e-action-pres-vfld-large-N}
    \eeq
For such vector fields, the LHS of the LE (\ref{e-loop-eqns})
identically vanish. Moreover, a vector field that solves
(\ref{e-action-pres-vfld-large-N})) will automatically solve
the finite-$N$ equation (\ref{e-action-pres-vfld-finite-N})),
though the converse need not be true. This is because all we
use is cyclicity and hermiticity of $G_I$, which is also true
of the $\F_I$. Now multiplying
(\ref{e-action-pres-vfld-finite-N}) by $\F_{K_1} \cdots
\F_{K_n}$, the same vector fields also satisfy
    \beq
    v_i^I S^{J_i i J_2} \F_{J_1 I J_2} \F_{K_1} \cdots \F_{K_n} = 0.
    \eeq
Taking expectation values, we see that symmetries of the action
in the large-$N$ limit automatically annihilate the change in
action term (with insertions) appearing in the finite-$N$ SDE
(\ref{e-SDE}). In particular, multiplying by $N^2$ and letting
$N \to \infty$, we see that the vector fields satisfying
(\ref{e-action-pres-vfld-large-N}) also annihilate the change
of action term on the LHS of the ${\cal O}(1/N^2)$ SDE
(\ref{e-subleading-SDE})
    \beq
    v_i^I S^{J_i i J_2} G^{(2)}_{J_1 I J_2 K_1 \cdots K_n} = 0.
    \eeq

Combining this with our result from section
(\ref{s-vol-pres-vfld-ann-terms-in-LE-and-SDE}) on volume
preserving vector fields, we come to the following conclusion.
Suppose the vector field $v$ is such that both LHS and RHS of
the large-$N$ LE identically vanish, $v_i^I S^{J_1 i J_2}
G_{J_1 I J_2} = 0 = v^I_i \eta^i_I$. Then the change in action
and change in measure term in the ${\cal O}(1/N^2)$ SDE also
vanish identically $v_i^I S^{J_i i J_2} G^{(2)}_{J_1 I J_2 K_1
\cdots K_n} = 0 = v_i^I \d_I^{I_1 i I_2} G^{(2)}_{I_1; I_2;
K_1; \cdots ;K_n}$. As a consequence, for such vector fields
(non-anomalous vector fields), the ${\cal O}(1/N^2)$ SDE become
WI (\ref{e-ward-id}) which may be summarized as $L_v G_K = 0$
for all $K$. This completes the proof of validity of the WI.

%--------------------------------------------------------
\subsection{Non-anomalous symmetries of specific models}
\label{s-non-anomalous-symm-of-specific-models}
%--------------------------------------------------------

It is straightforward to see that non-anomalous vector fields
form a Lie sub-algebra of the infinite dimensional Lie algebra
of measure preserving vector fields (section
\ref{s-vol-pres-vfld-form-lie-alg}). For, $L_v S(G) =0$ and
$L_w S(G) =0$ implies that $[L_v,L_w] S(G) =0$. However, this
Lie algebra is not necessarily infinite dimensional and depends
on the action of the matrix model.

This brings us to the task of determining the non-anomalous
infinitesimal symmetries of specific matrix models. In looking
for measure preserving vector fields, recall (section
\ref{s-change-in-meas-homogeneous-vfld}) that we could break up
the problem into finding homogeneous measure preserving vector
fields of a given rank\footnote{A homogeneous vector field $v$
of rank $n$ is one whose components $v_i^I$ are non-vanishing
only for $|I|=n$. We call rank-$1$ vector fields linear
transformations, rank-$2$ vector fields quadratic changes of
variable and so on.}. The same strategy does not work in
general for symmetries of the action. However, if the action is
itself a homogeneous polynomial\footnote{Examples include the
Gaussian, Chern-Simons and Yang-Mills models.}, then
(\ref{e-action-pres-vfld-large-N}) does not mix vector fields
of different ranks. In that case, every solution to
(\ref{e-action-pres-vfld-large-N}) is a sum of homogeneous
solutions. More generally, the action may not be a homogeneous
polynomial, as for a Gaussian $+$ Yang-Mills model. In such
cases, not every solution of (\ref{e-action-pres-vfld-large-N})
is necessarily a sum of homogeneous solutions, though there may
still be large classes of homogeneous solutions. For this
reason, we begin by determining non-anomalous homogeneous
action-preserving vector fields of low rank.

A priori, it is not clear that there are {\em any} vector
fields that leave both action and measure invariant. Indeed,
for a $1$-matrix model (section \ref{s-single-matrix}) there
are none. We were pleasantly surprised to find not just linear
but also non-linear non-anomalous symmetries for several
interesting multi-matrix models. We begin with linear
non-anomalous symmetries in section \ref{s-linear-symm} and
give examples of non-linear non-anomalous symmetries in section
\ref{s-non-linear-symm}.

%----------------------------------------------------------
\subsection{Examples of linear non-anomalous symmetries}
\label{s-linear-symm}
%----------------------------------------------------------

We determine linear symmetries of both action and measure for
the Gaussian $\La$-matrix model, Chern-Simons $3$-matrix model,
Yang-Mills and Gaussian+YM $\La$-matrix models. The linear
non-anomalous symmetries of the Gaussian, CS $3$-matrix model,
YM $2$-matrix model and Gaussian+YM 2-matrix models form the
orthogonal Lie algebra with respect to the covariance matrix,
$sl_3({\bf R})$, $sl_2({\bf R})$ and $o(2)$ Lie algebras
respectively. Not every multi-matrix model has non-trivial
linear non-anomalous symmetries. The pure-quartic 2-matrix
model $\tr S(A) = \tr (A^4 + B^4)$ or the model studied by
Mehta\cite{Mehta:1981xt}, $\tr S(A) = \tr [c A_1 A_2 +
(g/4)(A_1^4 + A_2^4)]$ have no non-trivial linear action
preserving symmetries. Linear symmetries form a closed Lie
algebra among themselves (section
\ref{s-vol-pres-vfld-form-lie-alg}), so their Lie brackets
cannot be used to generate new symmetries.

%----------------------------------------------------------
\subsubsection{Linear symmetries of Gaussian}
\label{s-gaussian-linear}
%----------------------------------------------------------

The Gaussian $\La$-matrix model is defined by the action $\tr
S(A) = \tr \half C^{ij} A_i A_j$ where $C^{ij}$ is a positive
real symmetric `covariance' matrix. We seek infinitesimal
linear transformations $\d A_i = v_i^j A_j$ that leave the
action as well as measure invariant in the large $N$ limit. In
section \ref{s-charac-mean-pres-tr} we found that the measure
preserving transformations are the traceless ones $v^i_i=0$
forming the Lie algebra $sl_\La({\bf R})$. Here we find that
the vector fields $v_i^j$ that preserve both the action and
measure in the large-$N$ limit are those that satisfy $v^i_m +
v^j_k C^{ki} C_{jm} = 0$. This is the condition that $v_i^j$ be
an orthogonal transformation with respect to a metric given by
the covariance. In particular, for a unit covariance $C^{ij} =
\d^{ij}$, these are the antisymmetric matrices.

For the expectation value of the Gaussian action to be
invariant at $N=\infty$, we need
    \beq
    L_v S(G) = v_k^j C^{kl} G_{lj} = \half (\tl v^{jl} + \tl v^{lj}) G_{lj} = 0 ~~ \forall ~~ {\rm cyclic~~
    and~~ hermitian~~} G_{lj}.
    \eeq
We used $C^{kl}$ and its inverse $C_{lm}$ to raise and lower
indices, $\tl v^{jl} = v^j_k C^{kl}$, $\tl v^{jl} C_{lm} =
v^j_m$, $C^{kl} C_{lm} = \d^k_m$. The condition for a symmetry
of the action is that $\tl v$ be anti-symmetric
    \beq
    \tl v^{jl} + \tl v^{lj} =0.
    \eeq
If $\tl v$ is anti-symmetric, then $v$ is automatically
traceless. So action preserving linear transformations are
automatically measure preserving. To see this we first rewrite
antisymmetry of $\tl v$ as a condition on $v$ by lowering an
index $v^i_m + v^j_k C^{ki} C_{jm} = 0$. Taking the trace we
get $v^i_i + v^j_k C^{ki} C_{ji} = 0 $ which implies $ v^i_i
=0$. Thus, the non-anomalous linear symmetries of the Gaussian
are given by vector fields $v_i^j L^i_j$ that are
anti-symmetric after raising an index with $C^{ik}$. In other
words, the orthogonal Lie algebra with respect to the metric
given by the covariance matrix. In particular, the dimension of
the space of linear symmetries $\half (\La^2 - \La)$, does not
change as we move around in the space of non-singular symmetric
covariance matrices.

{\fl \bf Example 1:} Consider a Gaussian two matrix model with
diagonal covariance $C^{ij} = diag(a,b)$. Then the condition
that $\d A_i = v_i^j A_j$ be action preserving in the large-$N$
limit is $v_1^1 = v_2^2 =0$ and $a v_1^2 + b v_2^1 = 0$. These
are the infinitesimal orthogonal transformations with respect
to the `metric' $diag(a,b)$. Such $v_i^j$ are traceless and
thus measure preserving as well.

{\fl \bf Example 2:} If the covariance of a Gaussian
$\La$-matrix model is a multiple of the identity, then the
action preserving transformations are the anti-symmetric
matrices ($v_i^j + v_j^i = 0$) which form the orthogonal Lie
algebra with respect to the metric $\d^{ij}$. Such matrices are
clearly traceless so that the non-anomalous linear symmetries
form the Lie algebra $o(\La)$.

%----------------------------------------------------------
\subsubsection{Linear symmetries of Chern Simons model}
\label{s-CS-linear}
%----------------------------------------------------------

The Chern-Simons $3$-matrix model has action
    \beq
    \tr S(A) = {2i \ka \over 3} \eps^{ijk} A_i A_j A_k = 2i \ka \tr A_1 [A_2,
        A_3] = 2 i \ka \tr (A_{123} - A_{132}).
    \eeq
So the coupling tensors are $S^{ijk} = {2 i \ka \over 3}
\eps^{ijk}$. We seek all vector fields $L_v = v^I_i L^i_I$ for
which the change in ($\ov{N} \times$) the action vanishes for
$N=\infty$ (second equality requires relabeling of indices)
    \beq
    L_v S^J G_J = v_i^I S^{J_1 i J_2} G_{J_1 I
    J_2} = {2 i \ka} v_i^I \eps^{ijk} G_{I j k} = 0.
    \eeq
$v_i^I$ are real and have no symmetry in $I$. They must satisfy
$v_i^I \eps^{ijk} G_{Ijk} = 0$ for all cyclic and hermitian
$G_K$. Specializing to linear transformations $A_i \to A_i +
v_i^j A_j$, they must satisfy
    \beq
    \sum_{1 \leq i,j,k,l \leq 3} v_i^l \eps^{ijk} G_{ljk} = 0.
    \eeq
We could also arrive at this condition by making a linear
change of variables in the action
    \beq
    \d \N \tr S= \Ntr (v^{i}_{1}A_{i23}+v^{i}_{2}A_{1i3} +
    v^{i}_{3} A_{12i}-v^{i}_{1}A_{i32}-v^{i}_{2}A_{13i}-v^{i}_{3}A_{1i2})
    =0.
    \eeq
Writing out all the terms and using cyclicity of $G_K$ this
condition simplifies dramatically to
    \beq
    \sum_{i=1}^3 v_i^i ~~(G_{123} - G_{132}) = 0.
    \eeq
Taking real and imaginary parts\footnote{$v_i^I \in {\bf R}$.
Hermiticity \& cyclicity $\implies G_{123}^* = G_{132}$ which
$\implies \Re G_{123} = \Re G_{132}$ and $\Im G_{123} = - \Im
G_{132}$.} we get the single condition $v^i_i ~\Im ~G_{123} =
0$, which must be satisfied for all $\Im~ G_{123}$. We conclude
that $v_i^j$ preserves the CS action iff it is traceless $v_i^i
= 0$. We recall (section \ref{s-charac-mean-pres-tr}) that
traceless linear transformations also preserve the matrix model
measure. Thus, the CS model has a maximal family of linear
non-anomalous symmetries.

From section \ref{s-vol-pres-vfld-form-lie-alg} we know that
the space of traceless real $v_i^j$ is the Lie algebra
$sl_3({\bf R})$, an $8$ dimensional space. The free parameters
can be chosen as $v^2_1, v^3_1,
v^1_2,v^3_{2},v^{1}_{3},v^{2}_{3},v^{1}_{1}$ and $v^{2}_{2}$
with $v^3_3 = -v^1_1 - v^2_2$. The corresponding symmetries are
an 8-parameter family of vector fields
    \beq
    L_v = v^2_1 L^1_2 + v^3_1 L^1_3 + v_2^1 L_1^2 + v_2^3 L_3^2 +
    v_3^1 L_1^3 + v_3^2 L_2^3 + v_1^1 L_1^1 + v_2^2 L_2^2 - (v_1^1 +
        v_2^2) L_3^3.
    \label{e-cs-linear-symm}
    \eeq

%----------------------------------------------------------
\subsubsection{Linear Symmetries of Yang-Mills model}
\label{s-ym-linear}
%----------------------------------------------------------

For 2 or more matrices and a real symmetric invertible metric
$g_{ij}$, the YM model has action $\tr S(A) = -\ov{4 \a} \tr
[A_i,A_j] [A_k,A_l] g^{ik} g^{jl}$. The expectation value of
the change in the action under a linear transformation $\d A_i
= v_i^j A_j $ in the large-$N$ limit can be written as
    \beq
    L_v S(G) = -\ov{\a} G_{jklm} \bigg(v_i^m g^{ik} g^{jl} - v_i^k g^{il} g^{jm}
    - v_i^j g^{im} g^{kl} + v_i^k g^{im} g^{jl} \bigg)
    \eeq
To identify symmetries of the action, we must select
independent $G_{jklm}$ and set their coefficients to zero.
First we restrict the sum to words $jklm$ up to cyclic
symmetry. Thus $\d S = 0$ iff
    \beq
    -\ov{\a} \sum_{jklm/cyc} G_{jklm} R^{jklm}   = 0
    \eeq
where the cyclically symmetric tensor $R^{jklm}$ is
    \beq
    R^{jklm} &=&  \bigg[ v_i^j (2 g^{il} g^{km} - g^{im} g^{kl} - g^{ik}
        g^{lm}) + cyclic (j \to k \to l \to m \to j) \bigg] \cr
    &=& (\tl v^{(jl)} g^{km} + \tl v^{(km)} g^{jl})
        - \half \bigg[ \tl v^{(jm)} g^{kl} + \tl v^{(jk)} g^{lm}
        + \tl v^{(kl)} g^{mj} + \tl v^{(lm)} g^{jk} \bigg].
    \eeq
Here we have used the metric to raise and lower indices $\tl
v^{jl} = v^j_i g^{il}$ and $v^j_m = \tl v^{jl} g_{lm}$ and
denoted the symmetric projection by $\tl v^{(jk)} = \half (\tl
v^{jk} + \tl v^{kj})$. We have still to account for the
hermiticity relations $\Re G_{jklm} = \Re G_{\overline{jklm}}$,
$\Im G_{jklm} = -\Im G_{\overline{jklm}}$. Now, $v_i^j$ and
$g^{kl}$ are real and $\Re G_{jklm}$ and $\Im G_{jklm}$ are
independent of each other, so we have $L_v S(G) = 0$ iff
    \beq
    \sum_{jklm/cyc} R^{jklm} \Re G_{jklm}  =0 {~~~ \rm and ~~~}
    \sum_{jklm/cyc} R^{jklm} \Im G_{jklm}  =0.
    \eeq
Now we must collect the coefficients of $\Re G_{jklm}$ and $\Re
G_{\overline{jklm}}$ and similarly for the imaginary parts and
restrict the sum to avoid $\overline{jklm}$ if $jklm$ has
already appeared. Two possibilities arise: either
$\overline{jklm}$ may be obtained from $jklm$ via cyclic
permutations or not. In the former case, $\Im G_{jklm}$
vanishes and the coefficient of $\Re G_{jklm}$ must vanish for
$v$ to be a symmetry of the action. Thus we get $R^{jklm} = 0$
if $jklm$ is cyclically related to $\overline{jklm}$. On the
other hand, if $jklm$ is not cyclically related to
$\overline{jklm}$, then collecting coefficients we have
    \beq
    \sum_{jklm/cyc, revers} \Re G_{jklm} (R^{jklm} +
    R^{\overline{jklm}}) = 0 ~~~~ {\rm and} ~~~
    \sum_{jklm/cyc, revers} \Im G_{jklm} (R^{jklm} -
    R^{\overline{jklm}}) = 0
    \eeq
Now the sums are over truly independent moments. Setting
coefficients to zero we get the pair of conditions $R^{jklm} +
R^{\overline{jklm}} =0$ and $R^{jklm} - R^{\overline{jklm}}
=0$, whose simultaneous solution is again $R^{jklm} =0$. We
conclude that the necessary and sufficient conditions for
$v_i^j$ to be a symmetry of the action are $R^{jklm} =0$. By
contracting with the non-singular metric to get a scalar,
    \beq
    R^{jklm} g_{jl} g_{km} = (6\La -4) v_i^i.
    \eeq
Since $\La \ne 2/3$, if $v$ is action preserving
($R^{jklm}=0$), then $\tr v=0$ and $v$ is automatically measure
preserving. Thus, non-anomalous linear symmetries of the
Yang-Mills model in the large-$N$ limit are characterized by
those $v$ for which the tensor $R^{jklm}$ vanishes. It suffices
to check this condition for each word $jklm$ up to cyclic
permutations and order reversals. Since $R^{jklm}$ depends only
on the symmetric projection of $\tl v^{jk}$, the anti-symmetric
part of $\tl v^{jk}$ is unconstrained! Thus, a sufficient
condition for $\tl v^{ij}$ to be a non-anomalous symmetry is
that it be anti-symmetric. However, this is not a necessary
condition; there are traceless\footnote{$\tr v = v^j_j = \tl
v^{jl} g_{lj}$.} $\tl v^{ij}$ with non-trivial symmetric
projections for which $R^{jklm} = 0$.

{\bf Example:} We will demonstrate this using the simplest
non-trivial example, the $2$-matrix Yang-Mills model with flat
metric $g_{ij} = \d_{ij}$. In this case, the action reads $\tr
S(A) = -\ov{2\a} \tr [A_1,A_2]^2$. Then $v^i_j = \tl v^{ik}
\d_{kj}$. The antisymmetric part of $\tl v$ automatically
satisfies $R^{jklm}=0$, so let us suppose that $\tl v^{ij}$ is
a traceless symmetric tensor, i.e. $\tl v^{(ij)} = \tl v^{ij}$
and $\tl v^{11} + \tl v^{22} = 0$. Then the six independent
components of $R^{jklm}$ are all identically zero
    \beq
    R^{1111} = 2(\tl v^{11} + \tl v^{11}) - 4 \tl v^{11} =0, &&
    R^{2222} = 2(\tl v^{22} + \tl v^{22}) - 4 \tl v^{22} =0, \cr
    R^{1112} = 2 \tl v^{12} - (\tl v^{12} + \tl v^{12}) =0, &&
    R^{1122} = -(\tl v^{11} + \tl v^{22}) =0, \cr
    R^{1212} = 2(\tl v^{11} + \tl v^{22}) =0, &&
    R^{1222} = 2 \tl v^{12} - (\tl v^{12} + \tl v^{12}) =0.
    \eeq
So {\em every} symmetric traceless $\tl v^{jk}$ satisfies
$R^{jklm}=0$. We conclude that for $\La=2$ and $g_{ij} =
\d_{ij}$, the Lie algebra of non-anomalous symmetries is
$sl_2({\bf R})$.

%----------------------------------------------
\subsubsection{Linear symmetries of Gaussian $+$ Yang-Mills}
\label{s-gauss-plus-ym-linear}
%----------------------------------------------

For $\La \geq 2$ let us consider a Gaussian $+$ Yang-Mills
matrix model with action
    \beq
    \tr S(A) = \half C^{ij} \tr A_i A_j - \ov{4\a} \tr [A_i,A_j][A_k,A_l] g^{ik}
    g^{jl}.
    \eeq
The simplest case which we will focus on is the two matrix
model with flat metric $g_{ij} = \d_{ij}$ and with covariance a
multiple of the identity $C^{ij} = m^2 \d^{ij}$. In this case
the action reads
    \beq
    \tr S(A) = \tr {m^2 \over 2} (A_1^2 + A_2^2) - {1 \over 2 \a} \tr
    [A_1,A_2]^2.
    \label{e-gauss-plus-ym-2-mat-action}
    \eeq
We know (sections \ref{s-gaussian-linear}, \ref{s-ym-linear})
that linear non-anomalous symmetries of the Gaussian and
Yang-Mills parts constitute the $o(2)$ and $sl_2({\bf R})$ Lie
algebras respectively. Their intersection is $o(2)$, which is
automatically a non-anomalous symmetry algebra of
(\ref{e-gauss-plus-ym-2-mat-action}). But these must be all the
linear symmetries, since there can be no cancelation between
$L_v S_{gauss}(G)$ which involves two point correlations and
$L_v S_{YM}(G)$ which involves $4$-point correlations
exclusively. The corresponding conclusion for $\La$ matrix
models (again with $C^{ij}$ a multiple of identity and $g_{ij}
= \d_{ij}$) is that the non-anomalous linear symmetries form
the orthogonal Lie algebra $o(\La)$.

%----------------------------------------------------------
\subsection{Examples of non-linear non-anomalous symmetries}
\label{s-non-linear-symm}
%----------------------------------------------------------

We exhibit homogeneous quadratic infinitesimal changes of
variable $\d A_i = v_i^{jk} A_j A_k$ which leave both action
and measure invariant in the large-$N$ limit. In particular, we
consider the $2$-matrix Gaussian with unit covariance, the
$3$-matrix Chern Simons model, the $2$-matrix
commutator-squared Yang-Mills model and the 2-matrix
Gaussian+YM model. We find a $2, 18, 6$ and $2$ dimensional
family of quadratic non-anomalous symmetries in these cases.
Moreover, we show that quadratic symmetries do not form a Lie
algebra by themselves. We demonstrate how to obtain non-trivial
non-anomalous {\em cubic} symmetries via their Lie brackets.

%----------------------------------------------------------
\subsubsection{Quadratic symmetries of the Gaussian model}
\label{s-quad-symm-gaussian}
%----------------------------------------------------------

Under an infinitesimal quadratic change of variable $\delta
A_{i}=v^{jk}_{i}A_{jk}$, the change in action of a Gaussian
model with unit covariance is
    \beq
    \d S = \tr \d^{ij} \d A_{i}A_{j}
    = \tr \d^{ij} v^{mn}_i A_{mnj}
    = \tr v^{mn}_i A_{mni}.
    \eeq
Specializing to a $2$-matrix model in the large-$N$ limit and
taking expectation values,
    \beq
    L_v S(G) =  v^{11}_{1} G_{111} + (v^{11}_{2}+v^{12}_{1}+v^{21}_{1}) G_{112} +
    (v^{12}_{2}+v^{12}_{2}+v^{21}_{2})G_{221}+v^{22}_{2}G_{222}.
    \eeq
where we have collected the coefficients of the four
independent third rank moments, which are all real after
accounting for cyclicity and hermiticity. Thus $L_v S=0$
implies
    \beq
    v^{11}_{1} = v^{22}_{2}=0, ~~~
    v^{11}_{2} = -v^{12}_{1}-v^{21}_{1}, ~~~
    v^{22}_{1} = -v^{12}_{2}-v^{21}_{2}.
    \eeq
To be non-anomalous, $v$ must be volume preserving as well:
$v^{ij}_{i}+v^{ji}_{i}=0$, which implies
(\ref{e-2-mat-quad-meas-pres-vfld}).
%    \beq
%    2 v^{11}_{1} + v^{21}_{2} + v^{12}_{2} =0, ~~~
%    2 v^{22}_{2}  + v^{12}_{1} + v^{21}_{1} =0.
%    \eeq
The solution of this system of linear equations is a two
parameter family
    \beq
    v^{11}_{1} = v^{22}_{2}=v^{11}_{2}=v^{22}_{1}=0, ~~~
    v^{12}_{1} = -v^{21}_{1}=a, ~~~
    v^{12}_{2} = -v^{21}_{2}=b.
    \eeq
Thus, the non-anomalous quadratic symmetries of the Gaussian
model with unit covariance are
    \beq
    \delta A_{1} = a\left[A_{1}, A_{2} \right], ~~~~
    \delta A_{2} = b\left[A_{1}, A_{2} \right], ~~~~~ a,b ~\in~ {\bf
    R}.
    \eeq
They correspond to the vector fields $L_{u_{a,b}} = a (L_{12}^1
- L_{21}^1)  + b (L_{12}^2 - L_{21}^2)$. The Lie bracket of
$L_{u_{a,b}}$ and $L_{u_{c,d}}$ is not a quadratic vector
field. Rather, it is a cubic non-anomalous vector field
    \beq
    [L_{u_{a,b}},L_{u_{c,d}}] &=& (ad-bc) \bigg\{ [L_{12}^1, L_{12}^2] - [L^1_{12},L^2_{21}]
     - [L^1_{21}, L^2_{12}] + [L^1_{21},L^2_{21}] \bigg\} \cr
    &=& (ad-bc) (L^2_{122} - L^1_{112} - 2 L^2_{212} + 2 L^1_{121} +  L^2_{221} - L^1_{211}
    ).
    \eeq
It corresponds to the one parameter family of infinitesimal
changes of variable
    \beq
    \d A_1 = (ad-bc) [A_1,[A_2,A_1]], ~~~ \d A_2 = (ad-bc)
    [[A_1,A_2],A_2], ~~~~ ad-bc ~\in~ {\bf R}.
    \eeq
There could, of course, be more cubic non-anomalous symmetries
that do not arise as Lie brackets of quadratic symmetries. It
is satisfying that our point of view tells us something
interesting even about the Gaussian matrix model.

%--------------------------------------
\subsubsection{Quadratic symmetries of Chern-Simons}
\label{s-quad-symm-CS}
%--------------------------------------

For a homogeneous quadratic change of variable, the change in
the expectation value of the Chern-Simons action $\tr S(A) = {2
i \ka \over 3} \eps^{ijk} A_{ijk}$ in the large-$N$ limit is
    \beq
    L_v S(G) = 2 i \ka v_i^{lm} \eps^{jki} G_{lmjk}.
    \eeq
To account for cyclicity of $G_{lmjk}$ we cyclically symmetrize
the coefficient. Let
    \beq
    T^{lmjk} &=& v_i^{lm} \eps^{jki} + v_i^{kl} \eps^{mji} + v_i^{jk} \eps^{lmi}
    + v_i^{mj} \eps^{kli}. \cr
    {\rm Then,~~~~~}  L_v S(G) &=&  {(i \ka/2)} \sum_{lmjk/\sim} T^{lmjk} G_{lmjk}
    \eeq
where the sum is restricted equivalence classes of $lmjk$ under
cyclic permutations. Accounting for hermiticity, we get that
$L_v S(G) = 0$ iff
    \beq
    \sum_{lmjk/\sim'} (T^{lmjk} + T^{\overline{lmjk}}) \Re G_{lmjk} = 0;
    {\rm ~~~ and ~~~}
    \sum_{lmjk/\sim'} (T^{lmjk} - T^{\overline{lmjk}}) \Im G_{lmjk} = 0
    \eeq
where now the sums are further restricted modulo order
reversal. Now we may set the coefficients to zero and after
adding and subtracting we find
    \beq
    L_v S(G) = 0 ~~ \Leftrightarrow ~~ T^{lmjk} = 0
    \eeq
where the condition is imposed for all words $lmjk$ modulo
cyclic permutations. There are $c(n=4,\La=3) = 24$ such words
for a $3$-matrix model. For $9$ of these words (1111), (2222),
(3333), (1112), (1222), (2333), (2223), (3111) and (3331),
$T^{lmjk}$ identically vanishes. The equations $T^{lmjk} =0 $
for each of the remaining $24-9=15$ words are listed below. The
words are indicated in parenthesis to the left of the equations
    \beq
    (1212) ~~ v_3^{12} = v_3^{21} && (1123) ~~ v_1^{11} + v_3^{31} + v_2^{12} = 0
    ~~~~~~~~~~~~ (3312) ~~ v_3^{33} + v_2^{23} + v_1^{31} = 0 \cr
    (1122) ~~ v_3^{12} = v_3^{21} && (1132) ~~ v_1^{11} + v_3^{13} + v_2^{21} = 0
    ~~~~~~~~~~~~ (3321) ~~ v_3^{33} + v_1^{13} + v_2^{32} = 0 \cr
    (2323) ~~ v_1^{23} = v_1^{32} && (1213) ~~ v_3^{13} + v_2^{21} - v_2^{12} - v_3^{31} = 0
    ~~~~ (3132) ~~ v_2^{32} + v_1^{13} - v_1^{31} - v_2^{23} = 0 \cr
    (2233) ~~ v_1^{23} = v_1^{32} && (2231) ~~ v_2^{22} + v_1^{12} + v_3^{23} = 0  \cr
    (3131) ~~ v_2^{31} = v_2^{13} && (2213) ~~ v_2^{22} + v_1^{21} + v_3^{32} = 0  \cr
    (3311) ~~ v_2^{31} = v_2^{13} && (2321) ~~ v_1^{21} + v_3^{32} - v_3^{23} - v_1^{12} =
    0.
    \eeq
%    \beq %\mbox{}
%\(
%\begin{array}{cccc}
%(1123)&\;\;\; v^{11}_{1}+v^{12}_{2}+v^{31}_{3}=0 \;\;\;\;& (1332)&\;\;\; v^{13}_{1}+v^{32}_{2}+v^{33}_{3}=0 \\
%(1132)&\;\;\; v^{11}_{1}+v^{21}_{2}+v^{13}_{3}=0 \;\;\;\;& (3123)&\;\;\; v^{31}_{1}+v^{23}_{2}+v^{33}_{3}=0 \\
%(1223)&\;\;\; v^{12}_{1}+v^{22}_{2}+v^{23}_{3}=0 \;\;\;\; &(3332)&\;\;\; vacuous \\
%(1232)&\;\;\; -v^{23}_{3}-v^{12}_{1}+v^{21}_{1}+v^{32}_{3}=0 \;\;\;\;& (2323)&\;\;\; v^{23}_{1}-v^{32}_{1}=0 \\
%(2132)&\;\;\; v^{21}_{1}+v^{22}_{2}+v^{32}_{3}=0 \;\;\;\;& (2332)&\;\;\;  v^{23}_{1}-v^{32}_{1}=0\\
%(2223)&\;\;\; vacuous \;\;\;\;& (1131)&\;\;\; vacuous \\
%(1323)&\;\;\; v^{13}_{1}-v^{31}_{1}-v^{23}_{2}+v^{32}_{2}=0 \;\;\;\;&
%(1213)&\;\;\; -v^{12}_{2}+v^{21}_{2}+v^{13}_{3}-v^{31}_{3}=0 \\
%(1331)&\;\;\; v^{13}_{2}-v^{31}_{2}=0 \;\;\;\;& (1313)&\;\;\; -v^{13}_{2}+v^{31}_{2}=0 \\
%(3331)&\;\;\; vacuous \;\;\;\;& (1112)&\;\;\; vacuous \\
%(1212)&\;\;\; v^{12}_{3}-v^{21}_{3}=0 \;\;\;\;& (1221)&\;\;\; -v^{12}_{3}+v^{21}_{3}=0 \\
%(2221)&\;\;\; vacuous
%\end{array} %\)
% \mbox{}
%    \eeq
Equations for volume preserving vector fields are
$v^{ij}_{i}+v^{ji}_{i}=0$, or explicitly
    \beq
    v^{11}_{1}+v^{21}_{2}+v^{31}_{3}+v^{11}_{1}+v^{12}_{2}+v^{13}_{3}
    = 0, &&
    v^{12}_{1}+v^{22}_{2}+v^{32}_{3}+v^{21}_{1}+v^{22}_{2}+v^{23}_{3}
    =0 \cr
    {\rm and} && v^{13}_{1}+v^{23}_{2}+v^{33}_{3}+v^{31}_{1}+v^{32}_{2}+v^{33}_{3}
    =0.
    \eeq
We see that $(1123)+(1132)$, $(2231)+(2213)$ and
$(3321)+(3312)$ are equivalent to these equations. So quadratic
symmetries of the CS action are also volume preserving.
Moreover
    \beq
    (1132)-(1123)=(1213); ~~
    (2213)-(2231)=(2321); ~~
    (3321)-(3312)=(3132).
    \eeq
So equations $(1213)$, $(2321)$, $(3132)$ may be discarded. And
equations $(1122), (2233), (3311)$ are redundant. Thus we are
left with $15-6 = 9$ independent equations for $3^3 = 27$
unknowns $v_i^{lm}$, leaving an $18$ parameter family of
non-anomalous quadratic symmetries of the CS model.

%They may be parameterized by
%    \beq
%   v^{22}_{1},v^{11}_{2},v^{33}_{2},v^{11}_{3},v^{22}_{3},v^{33}_{1},\\
%   v^{23}_{1}, \;\;(determines\;\; v^{32}_{1})\\
%   v^{13}_{2}, \;\;(determines\;\; v^{31}_{2})\\
%   v^{12}_{3}, \;\;(determines\;\; v^{21}_{3})\\
%   v^{12}_{2},v^{31}_{3},v^{21}_{2}, \;\;(determine\;\; v^{13}_{3}, v^{11}_{1})\\
%   v^{12}_{1},v^{23}_{3},v^{21}_{1}, \;\;(determine\;\; v^{32}_{3}, v^{22}_{2})\\
%   v^{13}_{1},v^{23}_{2},v^{32}_{2}, \;\;(determine\;\; v^{31}_{1},
%   v^{33}_{3})
%    \eeq

%-----------------------------------------------------
\subsubsection{Quadratic symmetries of Yang-Mills model}
\label{s-quad-symm-ym}
%-----------------------------------------------------

The action of the two matrix Yang-Mills model may be written
$S=-{1 \over 2 \a} tr[A_{1}, A_{2}]^{2} = \ov{\a} \tr (A_{1122}
- A_{1212})$. Under an infinitesimal homogeneous quadratic
change of variables $\delta A_{i}=v^{jk}_{i}A_{jk}$ the change
in the action $-\ov{\a} tr\left\lbrace([\delta
A_{1},A_{2}]+[A_{1}, \delta A_{2}])[A_{1},A_{2}] \right\rbrace$
becomes, in the large-$N$ limit
    \beq
    -\a L_v S(G) &=& v_i^{jk} L_{jk}^i (G_{1212} - G_{1122}) = v_i^{jk}
        (\d_{1212}^{I_1 i I_2} G_{I_1 jk I_2} - \d_{1122}^{I_1 i I_2} G_{I_1 jk I_2})
    \cr &=& G_{11212} (2 v_1^{11} + v_2^{12} + v_2^{21}) -
        G_{11122} (2 v_1^{11} + v_2^{21} + v_2^{12})
    \cr && + G_{12122} (2 v_2^{22} + v_1^{12} + v_1^{21})
        - G_{11222} (2 v_2^{22} + v_1^{12} + v_1^{21}).
    \eeq
%    \beq
%\delta S&=&tr\left\lbrace([\delta A_{1},A_{2}]+[A_{1}, \delta A_{2}])[A_{1},A_{2}] \right\rbrace \\
%&=&\tr \left\lbrace v^{jk}_{1}(A_{jk2}-A_{2jk})(A_{12}-A_{21})
%+ v^{jk}_{2}(A_{1jk}-A_{jk1})(A_{12}-A_{21}) \right\rbrace \\
%&=& \tr \left\lbrace
%v^{jk}_{1}(2A_{jk212}-A_{jk221}-A_{jk122})+v^{jk}_{2}(2A_{jk121}-A_{jk211}-A_{jk112})\right\rbrace.
%    \eeq
Since the moments that appear are independent, we set the
coefficients to zero:
    \beq
    2v^{11}_{1}+v^{12}_{2}+v^{21}_{2} =0 {\rm ~~~and~~~}
    2v^{22}_{2}+v^{12}_{1}+v^{21}_{1} = 0.
    \eeq
But these conditions are identical to those for a quadratic
vector field to preserve the measure of a $2$-matrix model
(\ref{e-2-mat-quad-meas-pres-vfld}). Since the above two
equations are independent, we have a $2^3 - 2 = 6$ parameter
family of non-anomalous homogeneous quadratic symmetries of the
$2$-matrix Yang-Mills model given in
(\ref{e-2-mat-quad-meas-pres-lie-derivatives}).
%    \beq
%    L_v &=& v^{ij}_k L^k_{ij} = v^{11}_2 L^2_{11} + v^{22}_1 L^1_{22}
%    + v^{12}_1 \left(L^1_{12} - \half L^2_{22} \right)
%    + v^{21}_1 \left(L^1_{21}-  \half L^2_{22}
%    \right) \cr && + v^{12}_2 \left(L^2_{12}- \half L^1_{11} \right)
%    + v^{21}_2 \left(L^2_{21}- \half L^1_{11} \right)
%    \eeq
It is remarkable that every measure preserving linear and
quadratic vector field also preserves the action of the YM
$2$-matrix model and CS $3$-matrix model in the large-$N$
limit. We wonder if this continues to hold for higher rank
symmetries or more matrices.

{\fl \bf Cubic Symmetry:} The Lie bracket of two non-anomalous
rank-$2$ vector fields (if $\ne 0$) is a rank-$3$ non-anomalous
vector field (since they form a Lie algebra). This is a way of
generating new symmetries. Consider two quadratic symmetries of
the YM model
    \beq
    L_v = a L^2_{21}- {a \over 2} L^1_{11} {\rm ~~~and~~~}
    L_u = b L^1_{22}.
    \eeq
which correspond to the choices $v^{21}_2 = a, ~
v^{11}_1=-\frac{a}{2}$ while all other $v^{jk}_{i}$ vanish, and
$u^{22}_1=b$ and all~other $u^{jk}_i$ vanish. Their commutator
is
    \beq
    [L_{v},L_{u}] = ab\left\lbrace[L^{2}_{21},L^{1}_{22}]-\frac{1}{2}[L^{1}_{11},L^{1}_{22}]
    \right\rbrace
    ~~=~~ ab\left\lbrace
    L^{1}_{212}-L^{2}_{222}+\frac{3}{2}L^{1}_{221}+\frac{1}{2}L^{1}_{122}
    \right\rbrace.
    \eeq
The non-vanishing components of the resulting non-anomalous
cubic symmetry are
    \beq
    w^{212}_{1} = ab, ~~~
    w^{221}_{1} = \frac{3}{2}ab, ~~~
    w^{122}_{1} = \frac{1}{2}ab, {\rm ~~~and~~~ }
    w^{222}_{2} = -ab.
    \eeq
One can also check explicitly that this defines a simultaneous
symmetry of the action and the measure. For example, the change
in action is
    \beq
    \delta \tr S&=&ab \tr\left\lbrace\left([A_{212}+\frac{3}{2}A_{221}
    +\frac{1}{2}A_{122}, A_{2}]-[A_{1},A_{222}] \right)[A_{1},A_{2}]
    \right\rbrace  = 0.
%    \cr
%    &=& \frac{ab}{2} \tr\left\lbrace A_{212212}-A_{212221}+ A_{221212}
%    -A_{221221}+ A_{122212}-A_{122221}+ A_{222112}-A_{222121}\right\rbrace\nonumber\\
%    &=&0.
    \eeq
The conditions for a homogeneous cubic vector field to be
volume preserving are
    \beq
    w^{min}_i + w^{nim}_i = 0 {\rm ~~and~~}
    w^{imn}_i + w^{inm}_i + w^{mni}_i + w^{nmi}_i = 0,
    \eeq
and $w_i^{jkl}$ satisfy these conditions as well.

%----------------------------------------------------------
\subsubsection{Quadratic symmetries of 2-matrix Gaussian + YM
model} \label{s-quad-symm-gauss-plus-ym}
%----------------------------------------------------------

Having determined homogeneous quadratic symmetries of Gaussian
and YM models, we get those for Gaussian+YM model $\tr S(A) =
\tr [{m^2 \over 2} (A_1^2 + A_2^2) - \ov{2\a} [A_1,A_2]^2]$ by
taking their intersection. For, there can be no cancelation
between rank $3$ \& $5$ tensors from the action of a
homogeneous quadratic vector field on the Gaussian and YM
terms. Since every quadratic measure preserving vector field
also preserves the YM action, the intersection is the same
family $\d A_1 = a [A_1, A_2], ~~ \d A_2 = b [A_1,A_2], a,b \in
{\bf R}$ as for the Gaussian (section
\ref{s-quad-symm-gaussian}).

%----------------------------------------------------------
\section{Supplementing loop equations with Ward identities}
\label{s-ward-id}
%----------------------------------------------------------

%----------------------------------------------------------
\subsection{Gaussian}
\label{s-WI-gauss}
%----------------------------------------------------------

LE of the Gaussian are not underdetermined. We do not need the
WI in this case. Nevertheless, the Gaussian does have
non-anomalous symmetries (sections \ref{s-gaussian-linear},
\ref{s-quad-symm-gaussian}), which lead to non-trivial WI. The
LE along with WI are an overdetermined system in this case.
Nevertheless, the WI are consistent with the LE and there is no
contradiction, as we have shown in section
\ref{s-sde-and-ward-id}. To illustrate this, consider a
$\La$-matrix model with unit covariance $\tr S(A) = \half
\d^{ij} A_i A_j$. The unique solution to the LE states that the
odd rank correlations vanish and the even ones $G_K$ are
determined by a planar version of Wick's theorem involving a
sum over non-crossing partitions of $K$ into pairs of indices.
For example $G_{ij} = \d_{ij}, ~~ G_{ijkl} = \d_{ij} \d_{kl} +
\d_{il} \d_{jk}$  etc. The WI corresponding to linear
non-anomalous symmetries are ($v^k_j$ are anti-symmetric)
    \beq
    L_{v}G_{K} = v^{k}_{j}L^{j}_{k}G_{K}
    ~=~ \sum_{i=1}^n v^{k}_{j}\delta^{j}_{k_{i}}G_{k_{1}\ldots k_{i-1} k k_{i+1}\ldots k_{n}}
    ~=~ \sum_{i=1}^n v^{k}_{k_{i}}G_{k_{1}\ldots k_{i-1} k k_{i+1}\ldots k_{n}} =
    0.
    \eeq
Thus, the moments must be $o(\Lambda)$ invariant tensors. We
can check that these WI are consistent with the LE. For
example, with $n=1$ we get the WI $v_l^k G_k =0$, for all
antisymmetric $v_l^k$. But there are anti-symmetric $v_k^j$
with non-vanishing determinant, and $G_j$ must lie in their
kernel which is trivial. So $G_j =0$, as implied by the LE. The
WI for odd $n$ are trivially satisfied by solutions to the LE,
since odd rank moments vanish. WI for $n=2$ are
    \beq
    v_{k_1}^k G_{k k_2} + v_{k_2}^k G_{k_1 k} = 0, ~~~~ \forall {\rm ~~~~antisymmetric~~~}
    v_l^k.
    \eeq
For $G_{ij} = \d_{ij}$, the LHS becomes $v_{k_1}^{k_2} +
v_{k_2}^{k_1}$ which vanishes on account of anti-symmetry of
$v$. Similarly we can check that the WI are consistent with the
LE for $n=4,6, \cdots$.

We could do the same for quadratic symmetries. Let us consider
the two-matrix Gaussian model. The WI $L_v G_K =0$ following
from quadratic non-anomalous symmetries (section
\ref{s-quad-symm-gaussian}) $L_v = a(L^1_{12} - L^1_{21}) +
b(L^2_{12} - L^2_{21})$ for arbitrary real $a, b$ are
    \beq
    \d_K^{L 1 M} G_{L 12 M} = 0 {\rm ~~~and ~~~}  \d_K^{L 2 M} G_{L 21 M} =
    0.
    \eeq
These WI are consistent with the LE (for $|K| \leq 4$, that we
checked, these WI are consequences of cyclicity and do not
contain new information). For more nontrivial use of the WI we
must progress to non-Gaussian multi-matrix models whose LE are
underdetermined.

%----------------------------------------------------------
\subsection{Gaussian plus Yang-Mills}
\label{s-WI-gauss-plus-ym}
%----------------------------------------------------------

The matrix integrals for correlations of the $2$-matrix
Gaussian + YM model, whose action is
    \beq
    \tr S(A) = \tr [ \frac{m}{2}(A^2_1 + A^2_2) - \frac{1}{2\a}
        [A_1,A_2]^2 ],
    \eeq
converge. Recall that the commutator of hermitian matrices is
anti-hermitian, and the square of an anti-hermitian matrix is
non-positive. Thus, the quartic term is non-negative. The
quadratic term ensures that as any matrix element goes to $\pm
\infty$, the action goes to $+\infty$. Thus, the Boltzmann
weight $e^{-N \tr S}$ vanishes at least exponentially fast as
any matrix element goes to $\pm \infty$. Thus, all polynomial
observables have finite expectation values. From this we
conclude that the LE and WI are rigorously valid. In section
\ref{s-underdeterminacy-of-LE} we obtained the LE for $|I| <
4$. They left a number of correlations undetermined. In section
\ref{s-gauss-plus-ym-linear} we found that linear non-anomalous
symmetries of this model form the $o(2)$ Lie algebra
parameterized by $v^i_j$ such that $v^1_1 = v^2_2=0$ and $v^1_2
= -v^2_1$. The corresponding WI, which we will use to
supplement the LE, read $TG_K=0$ for all words $K$, where
$T=L^2_1 -L^1_2$. These are listed in appendix
\ref{a-WI-list-gauss-plus-YM} for moments of rank up to $4$.
They imply that all $G_i$ vanish. The only $G_{ij}$ that might
be non-vanishing are $G_{11} = G_{22}$. All $3$-point $G_{ijk}$
vanish. $4$-point correlations vanish except possibly
$G_{1111}, G_{2222}, G_{1212}, G_{1122}$ and their cyclic
permutations. They must, however, satisfy the relations
$G_{1111} = G_{2222}$ and $G_{1111} = 2 G_{1122} + G_{1212}$.
Some of these conditions could also have been got from the LE,
(\ref{s-underdeterminacy-of-LE}). We need one more condition on
rank-$2$ moments and two more conditions on rank-$4$ moments to
determine all moments of rank $\leq 4$. The LE for $|I| =1$
gives one new condition
    \beq
    G_{11} = 1 + \ov{\a} (2 G_{1212} - 2 G_{1122}).
    \eeq
The LE for $|I|=2$ (section \ref{s-underdeterminacy-of-LE})
relate $3$ and $5$ point correlations.
%    \begin{itemize}
%\item $I=11$:
%    \beq
%    mG_{111}- \ov{\a} (2G_{11212}-2G_{11122})&=&2G_{1}=0 \implies G_{11122} = G_{11212}
%    \cr
%    mG_{112}- \ov{\a} (2G_{11121}-G_{11112}-G_{11211})&=& 0 \implies G_{112} = 0
%    \eeq
%
%\item $I=12$:
%    \beq
%    mG_{121}-\ov{\a}(2G_{12212}-G_{12221}-G_{12122})&=&G_{2}=0 \\
%    mG_{122}- \ov{\a} (2G_{12121}-G_{12112}-G_{12211})&=&G_{1}=0
%    \eeq
%
%\item $I=21$:
%    \beq
%    mG_{211}- \ov{\a} (2G_{21212}-G_{21221}-G_{21122})&=&G_{2}=0 \\
%    mG_{212}- \ov{\a} (2G_{21121}-G_{21112}-G_{21211})&=&G_{1}=0
%    \eeq
%
%\item $I=22$:
%    \beq
%    mG_{221}- \ov{\a} (2G_{22212}-G_{22221}-G_{22122})&=&0\Rightarrow G_{221}=0 \\
%    mG_{222}- \ov{\a} (2G_{22121}-2G_{22112})&=&2G_{2}=0
%    \eeq
%\end{itemize}
Using the fact that all $3$-point correlations vanish, they
tell us that $G_{11212} = G_{11122}$ and $G_{12122} =
G_{11222}$. Supplementing these LE with the WI for 5-point
correlations $T G_{ijklm} =0 ~\forall~ ijklm$ (which we do not
list explicitly), we are able to conclude that all rank-5
correlations vanish.

Thus far, we have found that the only correlations with rank
$\leq 5$ that could be non-vanishing are $G_{11}, G_{22},
G_{1111}, G_{2222}, G_{1212}$ and $G_{1122}$, up to cyclic
permutations. We have found $4$ relations among these $6$
unknowns:
    \beq
    G_{11} = G_{22}, G_{1111} = G_{2222}, G_{1111} = 2
    G_{1122} + G_{1212}, G_{11} = 1 + \ov{\a} (2 G_{1212} - 2
    G_{1122})
    \eeq
Thus, by use of the WI, we have reduced the underdeterminacy of
the LE. We could proceed further in this manner. The LE for
$|I|=3$ relate rank-$4$ and rank-$6$ moments, while the WI for
rank-$6$ moments give further conditions on rank-$6$ moments.
We could also look for additional conditions using the WI from
quadratic symmetries found in section
\ref{s-quad-symm-gauss-plus-ym}, but we postpone that. Our
purpose here was only to illustrate the general framework we
have developed. In a separate paper, we hope to return to a
more thorough study of the correlations of this model using the
LE and WI and their comparison with other approaches
\cite{eynard-kristjansen,kostov-3-color-problem} or monte-carlo
simulations\cite{gsk-gauss-ym-mc}.

\subsection{Chern-Simons 3-matrix model}
\label{s-WI-CS}
%----------------------------------------------------------

The CS 3-matrix model has action $\tr S(A) = {2i \ka \over 3}
\eps^{ijk} \tr A_{ijk} = 2 i \ka \tr A_1 [A_2,A_3]$. We expect
its matrix integrals to diverge. To see this, go to a basis
where $A_2$ is diagonal, then the action is independent of the
diagonal elements of $A_3$, due to the commutator. So
integration over the diagonal elements of $A_3$ would diverge.
Our derivation of the LE and WI holds at best formally for this
model. We do not know whether the LE and WI are a consistent
system for this action. Nevertheless, we consider them formally
to illustrate the general framework. We find no inconsistency,
at least for correlations up to rank $3$. The LE of the CS
3-matrix model are under determined (section
\ref{s-underdeterminacy-of-LE}). WI corresponding to
non-anomalous linear symmetries (\ref{e-cs-linear-symm}) are
obtained by imposing the conditions $L_v G_K =0$ for all $K$
and traceless $v^i_j$:
    \beq
    v_i^j \d_K^{K_1 i K_2} G_{K_1 j K_2} =0  ~~~ \forall
    ~~~    {\rm traceless~~~}  v^i_j.
    \eeq
These are the conditions that $G_K$ be (cyclic and hermitian)
invariant tensors of $\un{SL_3({\bf R})}$. We will work out the
WI explicitly for $|K| = 0,1,2,3$. For $K$ empty, this is a
vacuous condition, so put $K= k$ to get the WI $v_k^j G_j  = 0$
for all traceless $v_k^j$. There are traceless $v_k^j$ with
non-vanishing determinant, and $G_j$ must lie in the kernel of
such linear transformations. But this kernel is trivial, so
$G_j =0$. For $K=kl$ we get
    \beq
    v_l^j G_{kj} + v_k^j G_{jl} = 0 ~~~ \forall ~~ {\rm traceless}~~
    v.
    \eeq
Putting $k=l=1$ we get $v_1^1 G_{11} + v_1^2 G_{12} + v_1^3
G_{13} =0$. But this must hold for all real $v_1^1, v_1^2,
v_1^3$ so that $G_{11} = G_{12} = G_{13} = 0$. Putting $k=l=2$
we get $G_{21} = G_{22} = G_{23} =0$. Finally putting $k=l=3$
we get $v_3^1 G_{31} + v_3^2 G_{32} - (v_1^1 + v_2^2)$. Again
$v_3^1, v_3^2, v_1^1, v_2^2$ are freely specifiable so that
$G_{31} = G_{32} = G_{33} =0$. From this we conclude that all
$G_{ij} =0$. This is of course consistent with the remaining WI
gotten by putting $k=1,l=2$ etc since $G_{ij} =0$ is an obvious
solution of the homogeneous system $v_l^j G_{kj} + v_k^j G_{jl}
= 0$.

As for WI\footnote{$L^{i}_{i}$ (no sum over $i$) is a number
operator, it counts the number of $i$'s in a given moment. For
example, consider $G_{IiJiKiL}$, where none of the
multi-indices $I,J,K,L$ contain an $i$. Then
    \beq
    L^{i}_{i}G_{IiJiKiL}=G_{IiJiKiL}+G_{IiJiKiL}+G_{IiJiKiL}=3G_{IiJiKiL}.
    \eeq
The number operators commute $[L^{i}_{i},L^{j}_{j}]=0$. Thus
$\sum_i L^i_i$ measures the rank of a given moment. This can be
used to get most of the rank-$3$ moments of the CS model by
employing WI involving $L^1_1 -L^2_2$ and $L^1_1 - L^3_3$.} for
rank-3 correlations\footnote{Accounting for cyclicity and
hermiticity, the space of rank-3 tensors is $11$ dimensional,
see appendix \ref{a-cyclic-tensors-rank-n}.}, we set $K=klm$
and get
    \beq
    v_k^j G_{jlm} + v_l^j G_{kjm} + v^j_m G_{klj} = 0 ~~~
        \forall ~~ {\rm traceless}~~ v.
    \eeq
$k=l=m=1$ gives $v_1^1 G_{111} + v_1^2 G_{112} + v_1^3 G_{113}
=0$ whence $G_{111} = G_{112} = G_{113} = 0$. Similarly,
putting $k=l=m=2$ and $k=l=m=3$ we get $G_{122} = G_{222} =
G_{223} = G_{133} = G_{233} = G_{333} = 0$. The only remaining
undetermined rank-3 correlations are $G_{123}$ and $G_{132}$.
The remaining WI are either vacuous (e.g. $k=1,l=2,m=3$) on
account of $v$ being traceless or (e.g. $k=l=1, m=2$) give
$G_{123} + G_{132} = 0$.

To summarize, the WI due to linear non-anomalous symmetries
imply that all correlations of rank $\leq 3$ vanish except for
$G_{123}$ and $G_{132}$ (and their cyclic permutations), and
these are related by the WI $G_{123} + G_{132} = 0$. WI remedy
the underdeterminacy of LE of the CS model. The only
non-trivial condition from the LE was (see section 2.2.2 of
ref.\cite{gsk-loop-eqns})
    \beq
    \Im(G_{123} - G_{132}) = -{1 \over 2 \ka}
    \eeq
This, along with the WI $G_{123} + G_{132} = 0$ now allows us
to determine {\em all} correlations up to rank $3$, the only
non-vanishing ones (up to cyclic symmetry)
are\footnote{Hermiticity and cyclicity mean $G_{123}^* =
G_{132}$ which implies $\Re G_{123} = \Re G_{132}$ and $\Im
G_{123} = - \Im G_{132}$}
    \beq
    G_{123} = {1 \over 4i\ka}  {\rm ~~and~~}  G_{132} = -{1 \over 4i\ka}
    \eeq
We checked that this result is consistent with the WI
corresponding to quadratic non-anomalous symmetries obtained in
section \ref{s-quad-symm-CS}. So at least up to rank-$3$
moments, the WI cure the underdeterminacy problem of the LE! We
could proceed in this manner to higher rank correlations.

%----------------------------------------------------------
\subsection{$2$-matrix Yang-Mills: A cautionary tale}
\label{s-WI-ym}
%----------------------------------------------------------

The matrix integrals for correlations of the YM $2$-matrix
model $\tr S(A) = - \ov{2\a} \tr [A_1,A_2]^2$ do not converge
\cite{conv-YM-integrals} due to a similar argument as given for
the CS (section \ref{s-WI-CS}). Thus, our derivation of the WI
and LE is not strictly valid. We cannot be certain that they
form a consistent system. In fact, we find that the WI and LE
for this model do not form a consistent system when considering
rank-4 correlations. Despite several checks, we could find no
calculational error. We do not know the deeper reason for this
inconsistency, but suspect it could have something to do with
the lack of convergence of matrix integrals invalidating our
derivation of the WI and LE. Thus, it is probably good to be
cautious in formal use of the WI and LE.

The LE of the YM $2$-matrix model are underdetermined, (section
\ref{s-underdeterminacy-of-LE}). Recall that the LE do not
determine any moments of rank $1, 2$ or $3$. Here, the WI come
to the rescue. Recall (\ref{s-ym-linear}), that the
non-anomalous linear symmetries of this model form the Lie
algebra $sl_2({\bf R})$, spanned by $L^{1}_{1}-L^{2}_{2},
L^{1}_{2}, L^{2}_{1}$. The WI $L_v G_K =0$ for $|K| \leq 3$
suffice to determine all $1, 2,$ and $3$ point correlations,
and imply they are all zero.
%    \beq
%    L^{1}_{2}G_{1} = G_{2}=0, &&
%    L^{2}_{1}G_{2} = G_{1}=0, \cr
%    (L^{1}_{1}-L^{2}_{2})G_{11}= 2G_{11}=0 \Rightarrow G_{11}=0, &&
%    (L^{1}_{1}-L^{2}_{2})G_{22} = -2G_{22}=0 \Rightarrow G_{22}=0,
%    \cr
%    L^{1}_{2}G_{11} = 2G_{12}=0 \Rightarrow G_{12}=0, &&
%    (L^{1}_{1}-L^{2}_{2})G_{111} = 3G_{111}=0 \Rightarrow G_{111}=0,
%    \cr
%    (L^{1}_{1}-L^{2}_{2})G_{222} = -3G_{222}=0 \Rightarrow
%    G_{222}=0,
%    &&
%    (L^{1}_{1}-L^{2}_{2})G_{112} = G_{112}=0 \Rightarrow G_{112}=0,
%    \cr
%    L^{1}_{2}G_{112} = 2G_{122}=0 \Rightarrow G_{122}=0. &&
%    \eeq
Let us also consider WI for moments of rank 4:
    \beq
    (L^{1}_{1}-L^{2}_{2})G_{1111} = 4G_{1111}=0, &&
    (L^{1}_{1}-L^{2}_{2})G_{2222} = -4G_{2222}=0, \cr
    (L^{1}_{1}-L^{2}_{2})G_{1112} = 2G_{1112}=0, &&
    L^{1}_{2}G_{1112} = G_{1212}+2G_{1122}=0, \cr
    L^{1}_{2}G_{1212} = 2G_{2221}=0 \Rightarrow G_{2221}=0. &&
    \eeq
While the WI determine many rank-4 correlations, they give only
one relation between $G_{1212}$ and $G_{1122}$. To fix them we
use the simplest conditions coming from the first of the two LE
% (the second LE gives the same conditions)
    \beq
    I=1: ~~~~~
    2G_{1212}-2G_{1221}=-\alpha &;&
    I=2: ~~~~~    2G_{2212}-2G_{2221}=0.
    \eeq
The second equation is vacuous, but thanks to the WI we see
$G_{1222}=0$. As for the first equation, WI provides us with
the condition $G_{1212}+2G_{1122}=0$. So we get
    \beq
    G_{1212} = -\frac{\alpha}{3} {\rm ~~and~~}
    G_{1122} = \frac{\alpha}{6}.
    \eeq
However, this is not consistent with the WI for non-anomalous
quadratic symmetries (section \ref{s-quad-symm-ym}). In
particular the WI obtained from linear symmetries together with
    \beq
    (L^{1}_{12}-\frac{1}{2}L^{2}_{22})G_{112}=G_{1212}+\frac{1}{2}G_{1122}=0
    \eeq
implies $G_{1212}=G_{1122}=0$, which is a contradiction.

%---------------------------------
\section{Some outstanding questions}
\label{s-discussion}
%---------------------------------

A summary and discussion of the results of this paper was given
in the introduction. Here, we list some questions raised by our
work. (1) We have only addressed the exact determination of
normalized correlations in large-$N$ matrix models using the LE
and WI. But what about the partition function or free energy?
(2) It is interesting to know whether the LE and WI together
determine all single-trace correlations in the large-$N$ limit.
(3) We have only discussed infinitesimal non-anomalous
symmetries. Many models also possess discrete non-anomalous
symmetries, which lead to useful relations among correlations.
Some of these relations are actually a consequence of the LE or
WI. But in general, it may be necessary to supplement the LE
and WI by conditions from discrete symmetries. (4) Detailed
study of LE and WI of specific multi-matrix models should
clarify whether we need additional conditions to solve for all
correlations. (5) It is interesting to identify matrix models
with a maximal family of non-anomalous symmetries.
Interestingly, we found that the $3$-matrix CS model and the
$2$-matrix commutator-squared YM models each possesses a
maximal family of linear and quadratic non-anomalous
symmetries. (6) It is interesting to classify the solutions to
the simplest WI. For example, the correlations that satisfy WI
for linear symmetries of the Gauss+YM model must be invariant
cyclic hermitian tensors of the orthogonal Lie algebra. What is
the general form of such tensors?  (7) We observed that for $n
> 1$, Lie brackets of rank $n$ non-anomalous vector fields are
rank $n+1$ non-anomalous vector fields, provided they are
non-vanishing. It would be interesting to study this Lie
algebra of non-anomalous symmetries in specific examples. Can
it be infinite dimensional? If so, might the model be
integrable in some sense? (8) We wonder whether the full gauge
fixed Yang-Mills theory in the large-$N$ limit has any
additional non-anomalous symmetries besides Poincare invariance
and BRST invariance. Our work indicates that such symmetries
can be far from obvious and highly non-linear.

%------------------------------------
\section*{\normalsize Acknowledgements}
%------------------------------------

We thank Jean Yves Thibon for discussions on the dimension of
the space of cyclically symmetric tensors. We also thank S. G.
Rajeev, G. Arutyunov, S. Vandoren O. T. Turgut and G. 't Hooft
for discussions. GSK thanks the Feza Gursey Institute, Istanbul
for hospitality while a part of this work was done, and support
of the European Union in the form of a Marie Curie Fellowship.

%---------------------------------------------------------

\appendix

%----------------------------------------
\section{Alternative derivation of SDE preserving hermiticity}
\label{a-hermitian-derivation-of-SDE}
%----------------------------------------

Consider the change of variables corresponding to translation
by a constant Hermitian matrix $\epsilon$
    \beq
    (A_i)^a_b ~\to~ (A_i)^a_b + \eps^a_b, {\rm ~~~and~~~ }
    (A_j)^a_b ~\to~ (A_j)^a_b ~~ for ~~ j \neq i
    \eeq
in the integral ($[dA] = \prod_k d A_k$ is the Lebesgue measure
on independent matrix elements)
    \beq
    I=\int [dA] ~~(A_I)^a_b ~~ \F_K ~~ e^{-N tr S^J A_J}.
    \eeq
Here $\F_K = \Ntr A_K$. The value of the integral should be
unaltered under this change of variables\footnote{The following
formula is useful: $\frac{\pdr (A_I)^a_b}{\pdr (A_i)^c_d} =
\d_I^{I_1 i I_2}(A_{I_1})^a_e \d^d_f
    \d^e_c (A_{I_2})^f_b
    = \d_I^{I_1 i I_2} (A_{I_1})^a_c (A_{I_2})^d_b$.
Contracting $a$ with $b$ one gets $\frac{\pdr \tr A_I}{\pdr
(A_i)^c_d} = \d_I^{I_1 i I_2} (A_{I_1})^a_c (A_{I_2})^d_a
    = \delta_I^{I_1 i I_2} (A_{I_2 I_1})^d_c$.}
    \beq
    \d I &=& \eps^c_d \int [dA]~  \d_I^{I_1 i I_2}~ (A_{I_1})^a_c ~(A_{I_2})^d_b ~\F_K
    ~e^{-N \tr S^J A_J}  \cr
    && + ~~\eps^c_d \int [dA] ~ (A_I)^a_b ~ \d_K^{L i M} ~\N (A_{M L})^d_c ~ e^{-N \tr S^J
    A_J} \cr
    && - ~~ \eps^c_d \int [dA] ~ (A_I)^a_b ~\F_K~ N S^J
        \d_J^{J_1 i J_2} ~(A_{J_2 J_1})^d_c ~ e^{-N \tr S^J A_J} =0
    \eeq
where we also used the translation invariance of the measure.
Since this holds for arbitrary Hermitian $\epsilon$ we conclude
\footnote{It is shown in Adler's book \cite{adler} page. 26
that if $\tr(\eps H)$=0 for all Hermitian $\epsilon$ then
$H=0$}
    \beq
  \d_I^{I_1 i I_2} \bra (A_{I_1})^a_c (A_{I_2})^d_b \F_K \ket +
  \frac{1}{N} \d_K^{L i M} \bra (A_I)^a_b (A_{M L})^d_c  \ket
  =N S^{J_1 i J_2} \bra (A_I)^a_b \F_K (A_{J_2 J_1})^d_c \ket.
    \eeq
Contracting $a$ with $c$ and $b$ with $d$, and dividing both
sides by $N^2$ we get
    \beq
  \d_I^{I_1 i I_2} \bra \F_{I_1} \F_{I_2} \F_K \ket + \d_K^{L i M}  \Nsq \bra \F_{L I M}  \ket
      = S^{J_1 i J_2} \bra \F_{J_1 I J_2} \F_K \ket.
    \eeq
Since this must hold for every $I$ and $i$, it is equivalent to
the equations
    \beq
    v_i^I  \d_I^{I_1 i I_2} \bra \F_{I_1} \F_{I_2} \F_K \ket
    + v_i^I \d_K^{L i M}  \Nsq \bra \F_{L I M}  \ket
      = v_i^I S^{J_1 i J_2} \bra \F_{J_1 I J_2} \F_K \ket ~~~~
      \forall ~~~v_i^I ~\in~ {\bf R}.
    \eeq
In a completely analogous fashion, we repeat this calculation
with several insertions $\F_{K_1} \cdots \F_{K_n}$ and get the
Schwinger-Dyson equations obtained earlier in (\ref{e-SDE}).

%-----------------------------------------------------------
\subsection{Other possible changes of variables in matrix integrals}
\label{a-other-changes-of-var}
%-----------------------------------------------------------

LE are underdetermined, so do the $G_I$ satisfy other
equations? The WI {\em are} such equations and with the LE, may
go a long way towards fixing $G_I$. Here we consider two other
types of changes of variable in matrix integrals to see if they
give new equations. However, we do not find any.

Consider an infinitesimal change $\delta A_i=v_i^I A_I$ where
$v_i^I$ is a Hermitian matrix for each $I$ \& $i$; previously
they were real numbers. Conditions for invariance of partition
function are
    \beq
    -NS^{I_1 i I_2} \tr(A_{JI_2 I_1} v^J_i) +
    \d^{J_1 j J_2}_J \tr(v^J_j A_{J_1}) \tr A_{J_2}=0
    \eeq
%or
%    \beq
%    -NS^{I_{1}iI_{2}}(A_{JI_{2}I_{1}})^{a}_{b}(v^{J}_{i})^{b}_{a}
%    + \delta^{J_{1}jJ_{2}}_{J}(v^{J}_{j})^{b}_{a}(A_{J_{1}})^{a}_{b} \tr(A_{J_{2}})=0
%    \eeq
Since these must hold for arbitrary matrix elements
$[v_i^I]^a_b$ we get
    \beq
    -N S^{I_1 iI_2} (A_{J I_2 I_1})^a_b + \d^{J_1 j J_2}_J (A_{J_1})^a_b \tr A_{J_2}=0
    \eeq
But these are not equations for trace invariants. To get
equations for $G_I$, we must take a trace, divide by $N^2$ and
take expectation values. But this leads to the LE derived
before.

Next we consider an infinitesimal change $\delta A_i=[v_i^I,
A_I]$ with arbitrary hermitian matrices $v_i^I$. These types of
change of variable do not appear in BRST transformations but do
appear in gauge transformations $A_\mu \to A_\mu + \pdr_\mu \La
+ [A_\mu(x),\La(x)]$ . Invariance of partition function implies
    \beq
    \d^{I_1 j I_2}_I \tr (v^I_i A_{I_1}) \tr A_{I_2}
    - \d^{I_1 j I_2}_I \tr A_{I_1} \tr (v^I_i A_{I_2})
    + S^{I_1 j I_2} [A_{I_2 I_1}, A_I]=0.
    \eeq
Using the arbitrariness of $v_{i}^{I}$ we get
    \beq
    \d^{I_1 j I_2}_I (A_{I_1})^a_b \tr A_{I_2} -
    \d^{I_1 j I_2}_I \tr A_{I_1} (A_{I_2})^a_b =
    S^{I_1 j I_2 } [A_{I_2 I_1},A_I]^a_b.
    \eeq
As before, we must take a trace to get an equation for the
$G_I$, but in fact we get a triviality. Thus, we have not found
any equations for $G_I$ in addition to the LE and WI.

%------------------------------------------
\section{WI for Gaussian+YM model}
\label{a-WI-list-gauss-plus-YM}
%------------------------------------------

Below is a list of Ward identities $T G_K=0$ for moments of
rank up to $|K| = 4$ in the $2$-matrix Gaussian+YM model. They
correspond to the non-anomalous linear vector field $T = L_1^2
- L_2^1$.
    \beq
    TG_1  &=& -G_2 = 0; ~~~  TG_2 = G_1 = 0; ~~~
        TG_{11} = -G_{21} - G_{12} = -2G_{12}=0 \Rightarrow
        G_{12}=0;    \cr
    TG_{12} &=& G_{11}-G_{22}=0 \Rightarrow G_{11}=G_{22}; ~~~
        TG_{22} = 2G_{12}=0 \Rightarrow G_{12}=0; \cr
    TG_{111} &=& -3G_{112}=0\Rightarrow G_{112}=0; ~~~
        TG_{112} = G_{111}-2G_{122}=0\Rightarrow
        G_{111}=2G_{122}=0;  \cr
    TG_{122} &=& 2G_{112}-G_{222}=0\Rightarrow
        G_{222}=2G_{211}=0; ~~~
        TG_{222} = 3G_{122} \Rightarrow G_{122}=0; \cr
    TG_{1111} &=& -4G_{1112}=0; ~~~
        TG_{1112} = G_{1111}-2G_{1122}-G_{1212}=0\Rightarrow G_{1111}=2G_{1122} +
        G_{1212}; \cr
    TG_{1122} &=& -2G_{1222}+2G_{1112}=0 \Rightarrow
        G_{1112} = G_{1222}; ~~~
        TG_{1212} = 2G_{1112}-2G_{2221}=0; \cr
    TG_{1222} &=& 2G_{1122}+G_{1212}-G_{2222}=0; ~~~
        {\rm and~~} TG_{2222} = 4G_{1222}=0.
    \eeq

%----------------------------------------------------
\section{Is there a model for which Ward identities determine all $G_I$?}
\label{a-all-G_I-from-WI}
%----------------------------------------------------

Are there any non-trivial multi-matrix models where the WI are
sufficiently numerous to determine all (or a maximal set of)
correlations without using the LE? We seek models with a very
large family of non-anomalous symmetries. Trying to answer the
corresponding question in two dimensional quantum field theory
has proven very fruitful, as evidenced by the progress in 2d
conformal field theory. The latter are so symmetrical that a
maximal family of correlations can be determined by milking
conformal invariance. Here we make an elementary observation.
The WI $L_v G_K = 0$ are a system of homogeneous linear
equations. So they are either underdetermined (if the
determinant of the system vanishes) or admit only the trivial
solution $G_K = 0, ~\forall K$. Though the WI can give us much
information on correlations, they cannot determine all of them
except in the trivial case where they are all zero. For example
in a $2$-matrix model, if we consider an extreme (and probably
unrealistic) case where all measure preserving vector fields
are also action preserving\footnote{There may be no non-trivial
action with this property. We find the YM $2$-matrix model and
CS 3-matrix models are each such that every measure preserving
linear and quadratic symmetry is also action preserving.}, then
it follows that all correlations of rank up to $4$ vanish. This
leaves open the question of identifying non-trivial models with
a maximal family of non-anomalous symmetries, i.e. the ones for
which the WI are most useful.

\section{Cyclically symmetric tensors of rank $n$}
\label{a-cyclic-tensors-rank-n}
%---------------------------------------------------

The real dimension of the space of cyclic hermitian tensors
$G_{i_1 \cdots i_n}$ on a vector space $V$ of dimension $\La$
(i.e. $1 \leq i_1, \cdots, i_n \leq \La$) is
    \beq
    c(n,\La) = \ov{n} \sum_{d | n} \phi(d) \La^{n/d}
    \eeq
where $\phi(d)$ is Euler's totient
phi-function\footnote{$\phi(d)= $ number of positive integers
less than or equal to $d$ and coprime to $d$.} and the sum is
over all divisors of $n$. $c(n,\La)$ is the number of
independent correlations of rank-n in a $\La$ matrix model. We
thank the mathematician Jean Yves Thibon of Universit\'{e} de
Marne-la-Vall\'{e}e, France for sharing this formula with us.
This answers a question posed in Appendix A of
Ref.\cite{gsk-loop-eqns}. It comes from the character of
$GL(V)$
    \beq
    \ov{n} \sum_{d | n} ~\phi(d) ~~ p_d^{n/d}(\xi)  ~~~ {\rm where} ~~~
    p_d(\xi) = \sum_{i=1}^\La \xi_i^d {\rm ~~is ~the~ power~ sum~ symmetric~ function.}
     \eeq

\vs{-5}

%------------------------------------

%------------------------------------


\begin{thebibliography}{10}
%------------------------------------

\footnotesize

%\cite{'tHooft:1973jz}
\bibitem{thooft-large-N}
  G.~'t Hooft,
  ``A Planar Diagram Theory For Strong Interactions,''
  Nucl.\ Phys.\ B {\bf 72}, 461 (1974).
  %%CITATION = NUPHA,B72,461;%%

%\cite{Brezin:1977sv}
\bibitem{bipz}
E.~Brezin, C.~Itzykson, G.~Parisi and J.~B.~Zuber, ``Planar
Diagrams,'' Commun.\ Math.\ Phys.\  {\bf 59}, 35 (1978).
%%CITATION = CMPHA,59,35;%%


%\cite{Witten:1979kh}
\bibitem{witten-baryons-N}
  E.~Witten,
``Baryons In The 1/N Expansion,''
  Nucl.\ Phys.\ B {\bf 160}, 57 (1979).
  %%CITATION = NUPHA,B160,57;%%


%\cite{Makeenko:1979pb}
\bibitem{makeenko-migdal-eqn}
  Y.~M.~Makeenko and A.~A.~Migdal,
  ``Exact Equation For The Loop Average In Multicolor QCD,''
  Phys.\ Lett.\ B {\bf 88}, 135 (1979)
  [Erratum-ibid.\ B {\bf 89}, 437 (1980)];
  %%CITATION = PHLTA,B88,135;%%


%\cite{Cvitanovic:1980jz}
\bibitem{Cvitanovic:1980jz}
  P.~Cvitanovic,
  ``Planar Perturbation Expansion,''
  Phys.\ Lett.\ B {\bf 99}, 49 (1981).
  %%CITATION = PHLTA,B99,49;%%

%\cite{Yaffe:1981vf}
\bibitem{Yaffe}
  L.~G.~Yaffe,
  ``Large N Limits As Classical Mechanics,''
  Rev.\ Mod.\ Phys.\  {\bf 54}, 407 (1982).
  %%CITATION = RMPHA,54,407;%%


%\cite{Jevicki:1979mb}
\bibitem{Jevicki:1979mb}
  A.~Jevicki and B.~Sakita,
  ``The quantum collective field method and its application to the planar
  limit,''
  Nucl.\ Phys.\ B {\bf 165}, 511 (1980).
  %%CITATION = NUPHA,B165,511;%%

%\cite{Wadia:1980rb}
\bibitem{Wadia:1980rb}
  S.~R.~Wadia,
  ``On The Dyson-Schwinger Equations Approach To The Large N Limit: Model
  Systems And String Representation Of Yang-Mills Theory,''
  Phys.\ Rev.\ D {\bf 24}, 970 (1981).
  %%CITATION = PHRVA,D24,970;%%


\bibitem{voiculescu} D. Voiculescu, ``Lectures on free
probability theory'', Lectures on probability theory and
statistics, Ecole d'Eté de Probabilites de Saint-Flour XXVIII,
Lecture Notes in Math. 1738, 280-349, Springer, (1998).

%\cite{Douglas:1994zu}
\bibitem{douglas}
  M.~R.~Douglas,
 ``Large N gauge theory: Expansions and transitions,''
  Nucl.\ Phys.\ Proc.\ Suppl.\  {\bf 41}, 66 (1995)
  [arXiv:hep-th/9409098];
  %%CITATION = HEP-TH 9409098;%%
%\cite{Douglas:1994kw}
%\bibitem{Douglas:1994kw}
%  M.~R.~Douglas,
  ``Stochastic master fields,''
  Phys.\ Lett.\ B {\bf 344}, 117 (1995)
  [arXiv:hep-th/9411025].
  %%CITATION = HEP-TH 9411025;%%


%\cite{Gopakumar:1994iq}
\bibitem{gopakumar-gross}
  R.~Gopakumar and D.~J.~Gross,
  ``Mastering the master field,''
  Nucl.\ Phys.\ B {\bf 451}, 379 (1995)
  [arXiv:hep-th/9411021].
  %%CITATION = HEP-TH 9411021;%%

%\cite{Douglas:1994av}
\bibitem{douglas-li}
  M.~R.~Douglas and M.~Li,
``Free variables and the two matrix model,''
  Phys.\ Lett.\ B {\bf 348}, 360 (1995)
  [arXiv:hep-th/9412203].
  %%CITATION = HEP-TH 9412203;%%





%\cite{Akant:2001tu}
\bibitem{entropy-var-ppl}
  L.~Akant, G.~S.~Krishnaswami and S.~G.~Rajeev,
   ``Entropy of operator-valued random variables: A variational principle  for
  large N matrix models,''
  Int.\ J.\ Mod.\ Phys.\ A {\bf 17}, 2413 (2002)
  [arXiv:hep-th/0111263].
  %%CITATION = HEP-TH 0111263;%%




%\cite{Agarwal:2002cg}
\bibitem{hamiltonian-mat-mod-fisher-info}
  A.~Agarwal, L.~Akant, G.~S.~Krishnaswami and S.~G.~Rajeev,
``Collective potential for large-N Hamiltonian matrix models and
free  Fisher information,''
  Int.\ J.\ Mod.\ Phys.\ A {\bf 18}, 917 (2003)
  [arXiv:hep-th/0207200].
  %%CITATION = HEP-TH 0207200;%%


%\cite{Lee:1998ea}
\bibitem{lee-rajeev-closed}
  C.~W.~H.~Lee and S.~G.~Rajeev,
``A Lie algebra for closed strings, spin chains and gauge
theories,''
  J.\ Math.\ Phys.\  {\bf 39}, 5199 (1998)
  [arXiv:hep-th/9806002].
  %%CITATION = HEP-TH 9806002;%%



%\cite{Minahan:2002ve}
\bibitem{Minahan:2002ve}
  J.~A.~Minahan and K.~Zarembo,
  ``The Bethe-ansatz for N = 4 super Yang-Mills,''
  JHEP {\bf 0303}, 013 (2003)
  [arXiv:hep-th/0212208].
  %%CITATION = HEP-TH 0212208;%%




%\cite{Kazakov:2002yh}
\bibitem{kazakov-marshakov}
  V.~A.~Kazakov and A.~Marshakov,
``Complex curve of the two matrix model and its tau-function,''
  J.\ Phys.\ A {\bf 36}, 3107 (2003)
  [arXiv:hep-th/0211236].
  %%CITATION = HEP-TH 0211236;%%

%\cite{Kostov:1999xi}
\bibitem{Kostov:1999xi}
  I.~K.~Kostov,
  ``Conformal field theory techniques in random matrix models,''
  arXiv:hep-th/9907060.
  %%CITATION = HEP-TH 9907060;%%

%\cite{Eynard:2003kf}
\bibitem{eynard-loop-eqn-chain}
  B.~Eynard,
``Master loop equations, free energy and correlations for the
chain of matrices,''
  JHEP {\bf 0311}, 018 (2003)
  [arXiv:hep-th/0309036].
  %%CITATION = HEP-TH 0309036;%%


%\cite{Krishnaswami:2006ai}
\bibitem{gsk-loop-eqns}
  G.~S.~Krishnaswami,
``Multi-matrix loop equations: Algebraic and differential
structures and an approximation based on deformation
quantization,''
  JHEP {\bf 0608}, 035 (2006)
  [arXiv:hep-th/0606224].
  %%CITATION = HEP-TH 0606224;%%



%\cite{Green:1987sp}
\bibitem{green-schwarz-witten}
  M.~B.~Green, J.~H.~Schwarz and E.~Witten,
``Superstring Theory. Vol. 1: Introduction,'' Cambridge, Uk: Univ.
Pr. ( 1987) 469 P. ( Cambridge Monographs On Mathematical Physics).
%\href{http://www.slac.stanford.edu/spires/find/hep/www?irn=1755021}{SPIRES entry}

%\cite{Makeenko:2002uj}
\bibitem{makeenko-book}
Y.~Makeenko, ``Methods of contemporary gauge theory,''
Cambridge Univ. Pr. (2002).
%\href{http://www.slac.stanford.edu/spires/find/hep/www?irn=5355460}{SPIRES entry}


%\cite{Itzykson:1980rh}
\bibitem{itzykson-zuber-qft}
  C.~Itzykson and J.~B.~Zuber,
``Quantum Field Theory,'' McGraw-Hill (1980), republished by Dover
(2005).
%\href{http://www.slac.stanford.edu/spires/find/hep/www?irn=787043}{SPIRES entry}


%\cite{Mehta:1981xt}
\bibitem{Mehta:1981xt}
  M.~L.~Mehta,
``A Method Of Integration Over Matrix Variables,''
  Commun.\ Math.\ Phys.\  {\bf 79}, 327 (1981).
  %%CITATION = CMPHA,79,327;%%



% \bibitem{ym-cs-3-mat} hep-th/0608031

%\cite{Austing:2001bd}
\bibitem{conv-YM-integrals}
  P.~Austing and J.~F.~Wheater,
  ``The convergence of Yang-Mills integrals,''
  JHEP {\bf 0102}, 028 (2001)
  [arXiv:hep-th/0101071].
  %%CITATION = HEP-TH 0101071;%%



%\cite{Eynard:fn}
\bibitem{eynard-kristjansen}
B.~Eynard and C.~Kristjansen, ``An Iterative Solution Of The
Three-Colour Problem On A Random Lattice,'' Nucl.\ Phys.\ B
{\bf 516}, 529 (1998) [arXiv:cond-mat/9710199].
%%CITATION = COND-MAT 9710199;%%

%\cite{Kostov:2000vn}
\bibitem{kostov-3-color-problem}
I.~K.~Kostov, ``Exact solution of the three-color problem on a
random lattice,'' Phys.\ Lett.\ B {\bf 549}, 245 (2002)
[arXiv:hep-th/0005190].
%%CITATION = HEP-TH 0005190;%%

%\cite{Krishnaswami:2003pg}
\bibitem{gsk-gauss-ym-mc}
  G.~S.~Krishnaswami,
  ``Variational ansatz for gaussian + Yang-Mills two matrix model compared
  with Monte-Carlo simulations in 't Hooft limit,''
  arXiv:hep-th/0310110.
  %%CITATION = HEP-TH 0310110;%%

\bibitem{adler}
S. L. Adler, {\em Quantum Theory as an Emergent Phenomenon},
Cambridge University Press (2004).



\end{thebibliography}
\end{document}